 \documentclass[10pt,preprint]{aastex}  % per 1 colonna 

\shorttitle{Protostars in Vela-D cloud}

\shortauthors{Strafella et al.}

\begin{document}


\title{The YSO Population in the Vela-D Molecular Cloud.}


\author{F. Strafella\altaffilmark{1}, D. Lorenzetti\altaffilmark{2}, T. Giannini\altaffilmark{2}, 
        D. Elia\altaffilmark{3}, Y. Maruccia\altaffilmark{1}, B. Maiolo\altaffilmark{1},
        F. Massi\altaffilmark{4}, L. Olmi\altaffilmark{4}, S. Molinari\altaffilmark{3}, S. Pezzuto\altaffilmark{3} }

\altaffiltext{1}{Dipartimento di Matematica e Fisica, Universit\`a del Salento, I-73100 Lecce, Italy}
\email{francesco.strafella@le.infn.it}
\altaffiltext{2}{INAF-Osservatorio Astronomico di Roma, Via Frascati 33, I-00040 Monte Porzio, Italy}
\altaffiltext{3}{INAF-IAPS, via Fosso del Cavaliere 100, I-00133 Roma, Italy}
\altaffiltext{4}{INAF-Osservatorio di Arcetri, Largo E.Fermi 5, I-50125 Firenze, Italy}


\begin{abstract}
We investigate the young stellar population in the Vela Molecular Ridge, Cloud-D (VMR-D), a star forming  
(SF) region observed by both {\it Spitzer}/NASA and {\it Herschel}/ESA space telescope. 
The point source, band-merged, {\it Spitzer}-IRAC catalog complemented with MIPS photometry 
previously obtained is used to search for candidate young stellar objects (YSO), also including 
sources detected in less than four IRAC bands. Bona fide YSO are selected by using appropriate 
color-color and color-magnitude criteria aimed to exclude both Galatic and extragalactic contaminants. 
The derived star formation rate and efficiency are compared with the same quantities characterizing 
other SF clouds. Additional photometric data, spanning from the near-IR to the submillimeter, 
are used to evaluate both bolometric luminosity and temperature for 33 YSOs located in a region 
of the cloud observed by both {\it Spitzer} and {\it Herschel}. The luminosity-temperature diagram 
suggests that some of these sources are representative of Class 0 objects with bolometric temperatures 
below 70 K and luminosities of the order of the solar luminosity. 
Far IR  observations from the {\it Herschel}/Hi-GAL key project for a survey of the Galactic plane 
are also used to obtain a band-merged photometric catalog of {\it Herschel} sources aimed to 
independently search for protostars. We find 122 {\it Herschel} cores located on the molecular cloud, 
30 of which are protostellar and 92 starless. The global protostellar luminosity function is obtained 
by merging the {\it Spitzer} and {\it Herschel} protostars. Considering that 10 protostars are 
found in both {\it Spitzer} and {\it Herschel} list it follows that in the investigated region 
we find 53 protostars and that the {\it Spitzer} selected protostars account for approximately 
two-thirds of the total.
\end{abstract}


\keywords{Stars: formation --- Stars: protostars --- (Stars:) circumstellar matter --- catalogs}


%%%%%%%%%%%%%%%%%%%%%%%%%%%%%%%%%%%%%
\section{Introduction}

The Vela Molecular Ridge Cloud-D (hereafter VMR-D) 
($260\degr  \lesssim \ell \lesssim 264\degr$; $\left| b \right| \lesssim 1\degr$) 
is part of a giant molecular complex located along the Galactic plane 
\citep[$260\degr \lesssim \ell \lesssim 272\degr$; $\left| b\right| \lesssim 3\degr$][]{Mur91} 
and is then well suited to represent a typical star forming region (SFR) of our Galaxy. For this reason 
a subregion of this cloud has been the subject of many previous papers, dealing with different 
observational aspects of the star formation (SF) as, e.g., the presence of outflows \citep{Wou99,Eli07}, 
jets \citep{Lor02,Gia05,Gia13}, and clustering \citep{Mas00}. The continuum submillimeter emission in the 
VMR-D cloud was surveyed by \citet{Mas07} who catalogued 29 resolved dust cores and also obtained 
a further list of 26 unresolved candidate cores. More recently, thanks to the opportunity 
offered by the {\it Spitzer Space Telescope}, the VMR-D region was observed with 
the IRAC ($\lambda$ = 3.6, 4.5, 5.8, 8.0 $\mu$m) and MIPS ($\lambda$ = 24, 70 $\mu$m) focal 
plane instruments, obtaining in this way six mosaics, covering about 1.2 deg$^2$, 
that have been analyzed to produce a merged photometric {\it Spitzer}-IRAC point-source catalog 
(hereinafter {\it Spitzer}-PSC) 
complemented with MIPS photometry \citep[][hereinafter Paper I]{Str10}. Further observational progress 
was made when the BLAST experiment \citep{Pas08} mapped the whole Vela Molecular 
Ridge in the far-IR (FIR) spectral region ($\lambda$ = 250, 350, 500 $\mu$m), complementing in this way the 
{\it Spitzer} spectral coverage towards long wavelengths. These observations were discussed by \citet{Olm09} 
who obtained a catalog of dense cores/clumps in the VMR-D cloud. 
The last important observational progress was made with the {\it Herschel Space Observatory} which, 
surveying the Galactic plane in the framework of the Hi-GAL key project, partially mapped 
the VMR-D region in the FIR spectral range ($\lambda$ = 70, 160, 250, 350, 500~$\mu$m) with 
an almost double spatial resolution and sensitivity with respect to BLAST. 
Here we also analyze these observations for the first time, thanks to the support of the 
Hi-GAL collaboration that provided us with the corresponding calibrated maps. 
These have been used to extract five single band photometries that constitute another 
important spectral extension of our information about this region. 

Utilizing these observations as well as the photometric catalogs made 
available by the 2MASS and WISE all-sky surveys, we have the opportunity to study in much more detail 
the young stellar objects (YSOs) located in the VMR-D cloud. These objects can be 
efficiently identified through their infrared colors \citep[see, e.g.,][]{Har07, Gut09, Kry12} that 
we also exploit here to discriminate genuine YSOs from other contaminant populations of both Galactic 
and extragalactic origin.
All these observational data and tools give us the opportunity to investigate, with unprecedented 
sensitivity and accuracy, the characteristics of the SF in the VMR-D cloud. 

In Paper I, we already considered a subsample of 8796 sources, detected in all the four IRAC bands 
out of the $\approx$~170,000 sources listed in the {\it Spitzer}-PSC, to carry out a preliminary study 
of the YSO population in this cloud. Here, to obtain a more complete and accurate view, we 
revisit our previous preliminary census of the YSOs by including, besides the sources already 
considered in Paper I, all the other sources detected in at least two out of the four IRAC bands, 
with the additional requirement that sources detected in only two bands are also detected in 
the MIPS 24~$\mu$m band. 

As in Paper I, we adopt the usual classification of the YSOs based on the near and mid-IR spectral 
slope defined as $\alpha=d(\log{\lambda F_\lambda})/d(\log{\lambda})$, a parameter originally 
introduced by \citet{Lad87} to characterize the spectral energy distribution (hereinafter SED) 
of these objects that is still largely used and generally interpreted as a proxy of the 
evolutionary phases experienced by the YSOs. In this scheme, it is customary to distinguish four 
different classes of YSOs, according to the value of this slope in 2~$\lesssim \lambda(\mu$m)$\lesssim$~20 
spectral interval : Class I with $\alpha > 0.3$, flat spectrum (hereinafter FS) for $-0.3< \alpha < 0.3$, 
Class II for $-1.6 < \alpha < -0.3$, and finally Class III showing $\alpha < -1.6$. 
Given the evolutionary significance usually attributed to this parameter 
\citep[but see also][for alternative schemes]{Che95,You05}, 
Class I objects represent an early phase in the evolutionary path of the YSOs towards the ZAMS. 
Their spectral slope is produced by a central object that already attained 
its entire initial main-sequence mass, but is still surrounded by a remnant infall envelope and possibly 
an accretion disk. This description naturally implies the existence of an even earlier phase, the Class 0 
as originally suggested by \citet{And93}, in which the central object is at the end of the free-fall phase 
but is still increasing its mass by accreting material from a thick surrounding envelope. This fact makes 
the Class 0 phase difficult to detect in the near IR and then difficult to classify in the framework of the 
aforementioned scheme. 
In this observational classification both Class 0 and I are representative of early protostars 
still surrounded by a remnant infall envelope, although only in the Class 0 stage the envelope 
is presumed to be more massive than the central object and then opaque to near and possibly also 
to mid-IR wavelengths (both ranges hereinafter referred to as MIR). 

To study the SF mechanism in the VMR-D cloud we proceed along two approaches, both aimed to select 
candidate protostellar sources by exploiting all the information collected on the continuum emission. 
The first consists in selecting protostellar candidates by analyzing the {\it Spitzer} observations 
through a pipeline based on the criteria used by \citet{Har07} and \citet{Gut09}, that have been proven to 
be effective in isolating different kinds of possible contaminant sources. In the following, we shall refer 
to these sources as ``{\it Spitzer} selected'' and their analysis will be based on the SEDs assembled by 
collecting all the available photometries in what we shall call the MIR catalog.
The second approach adopts the FIR point of view because it is based on the analysis of the {\it Herschel} 
observations that, being carried out at longer wavelengths, are more sensitive to cold and 
young protostars. The sources selected in this way will be referred to as ``{\it Herschel} selected'' 
and their SEDs will be arranged in the FIR catalog resulting from the {\it Herschel} photometry 
complemented with all the available photometries at shorter and longer wavelengths. 
Merging the results obtained in these two ways we expect to obtain a more complete view of the 
protostellar population in VMR-D as well as new insights on the SF process in this molecular cloud. 

In the following, a brief account of the observational data obtained with {\it Spitzer} and {\it Herschel} 
is given in Section~\ref{Observations} along with a description of the procedures involved in data 
reduction and photometry. This section also illustrates how we obtain two multiwavelength source catalogs 
for the VMR-D region. The first, called MIR catalog, by complementing our previous {\it Spitzer}-PSC catalog 
with a set of additional observational data and the second, the FIR catalog, 
by assembling the five band {\it Herschel} photometry in a single band-merged list of sources.

In Section~\ref{Sources_selection} we select bona fide YSOs out of both catalogs by 
means of appropriate selection criteria and in Section~\ref{sources_classification} we discuss 
the census and classification of these sources. We also obtain a new estimate of the SF rate and 
efficiency that is more accurate than that obtained in Paper I based on the sole sources detected 
in all the four IRAC bands. 

In Section~\ref{Bolometric_quantities}, bolometric temperatures and luminosities are derived 
for all those sources for which a reasonably complete photometric information is available. 
Here the L$_\mathrm{bol}$ vs T$_\mathrm{bol}$ diagram is used to further characterize the 
YSOs and infer on their evolutionary status.

In Section~\ref{Discussion} we compare the SF rate and efficiency obtained for VMR-D 
with those obtained in other SFRs and discuss the global protostellar luminosity function (PLF). 
For those sources detected by {\it Herschel} at FIR wavelengths, the envelope mass has 
been also derived by exploiting a simple modified blackbody model. The luminosity-mass 
diagram is briefly discussed in the light of the protostellar evolutionary scenario. 

Finally, in Section~\ref{section_conclusions}, our conclusions are summarized.

%%%%%%%%%%%%%%%%%%%%%%%%%%%%%%%%%%%%
\section{Observational Data}\label{Observations}

The observational data on the continuum emission of the objects in VMR-D have been collected from 
many different sources including existing catalogs, public surveys, and new observations as in 
the case of the {\it Herschel} unpublished photometry. 
Dealing with a conspicuous amount of observational material spread over quite a large spectral 
interval, we decided to search for candidate YSOs exploiting all the spectral characteristics 
of their continuum emission. Our approach is twofold and corresponds to adopt a ``MIR point of view'' 
as well as a ``FIR point of view'' to identify protostellar candidates by means of procedures 
typically adopted in previous studies of SFRs observed with {\it Spitzer} and {\it Herschel}, 
respectively. 
In the following, to identify the sky regions involved in the different observations of 
the VMR-D cloud, we shall refer to Figure~\ref{fig_ds9_sources+coverage} that shows the 
8~$\mu$m IRAC image of the VMR-D cloud with overlayed the contours delimiting the areas 
observed by different instuments as well as the contour of the investigated cloud. 

\subsection{MIR catalog} \label{MIR_catalog}
We start considering the sources of the {\it Spitzer}-PSC catalog covering the region delimited 
by the cyan polygon in Figure~\ref{fig_ds9_sources+coverage}. To arrange all of the available 
continuum observations for an easy access and at the same time obtain the SEDs, we generated 
a new catalog containing the fluxes from all the available photometric surveys of this cloud 
from the near-IR to the submillimeter spectral range. 
This catalog is essentially a spectral extension of our previous {\it Spitzer}-PSC catalog and contains, besides the 
{\it Spitzer}-IRAC/MIPS (3.6, 4.5, 5.8, 8.0, 24, 70 $\mu$m) photometry, also 2MASS (J, H, K bands), 
WISE (3.4, 4.6, 12, and 22 $\mu$m), {\it Herschel} (70, 160, 250, 350, and 500~$\mu$m), and SEST-SIMBA 
(1200 $\mu$m) observations. 

%nnnnnnnnnnnn  associations:
Given the heterogeneous nature of the observational data, we adopted specific criteria 
in associating to the same source fluxes obtained at different wavelengths and with different 
spatial resolutions. This is a typical problem in assembling information coming 
from different catalogs and it is usually mitigated by requiring that the distance between 
the centroids of the sources to be associated is smaller than a predetermined value depending 
on the spatial resolution of the specific catalogs. 
This is also our approach even if more complex association procedures could be devised, involving 
further considerations on the flux values to be associated and, in turn, on the possible underlying 
spectral shapes \citep[see, e.g.,][]{Ros09, Bud08}. In our case we prefer to consider the positional 
uncertainty, instead of the beamsize, as a measure of the spatial accuracy of a given catalog 
because it is usually evaluated after an accurate statistical analysis and is generally more conservative. 
In this way we minimize possible mismatches in associating to the same source the fluxes from 
different catalogs, especially in crowded regions, even if we pay this choice with the risk 
to miss some source association. 
%nnnnnnnnnnnnn

Our procedure adopts the fiducial positions of the sources reported in the {\it Spitzer}-PSC catalog, 
in this way considering the other catalogs as ancillary, and are used to find possible 
counterparts at other wavelengths. The final product is a catalog containing the same 
number of entries as the {\it Spitzer}-PSC, but complemented with the associated fluxes observed at different 
wavelengths.
In Table~\ref{tab_bandcomplementing} we report the association radii adopted for the catalogs 
considered in this work, along with the associated beamsizes, completeness fluxes, and relevant 
references. 

In practice, we consider as acceptable the associations for which the distance $\Delta\theta$ 
between the position of the {\it Spitzer}-PSC source and the position of a source in a given catalog is 
less than the sum of the corresponding uncertainties, namely $\Delta\theta~<~\delta\theta_\mathrm{PSC} + \delta\theta_\mathrm{cat}$. 
In the case of multiple associations we always adopt the nearest source to the fiducial {\it Spitzer}-PSC position, 
allowing however that multiple {\it Spitzer}-PSC sources could be similarly associated to the same object of another catalog, 
a problem that will be taken into account later in the analysis. 
After scanning all the considered catalogs for positional association with the {\it Spitzer}-PSC sources we obtained 
a set of corresponding fluxes that, once assembled, constitute the MIR multiwavelength catalog of our interest.

\subsection{FIR catalog} \label{FIR_catalog}

The VMR-D cloud is located across the Galactic plane and it has been partially mapped 
by the {\it Herschel} space telescope during the completion of the Hi-GAL key project 
for a FIR survey of the Galactic plane \citep{Mol10}. As is shown in 
Figure~\ref{fig_ds9_sources+coverage} the survey coverage is incomplete because 
at these longitudes the observing strategy of the Hi-GAL survey preferred to follow 
the warp of the Galaxy instead of the Galactic plane.
The VMR-D cloud has been also mapped by the BLAST experiment at 250, 350 and 500~$\mu$m 
albeit with a significantly lower spatial resolution and sensitivity. 
Because of this, in the following we prefer to use the {\it Herschel} observations instead of 
the BLAST ones, alleviating in this way the problems related to both the strong and 
variable background and the source confusion. 

Thanks to the Hi-GAL consortium we obtained calibrated maps \citep{Ber00} of our region in which 
the artifacts introduced by the map-making technique have been removed or heavily attenuated by 
means of a weighted post-processing of the maps themselves \citep{Pia12}. 
The source detection and photometry has been carried out on these maps by using the CuTeX package 
\citep{Mol11}, a software specifically designed to detect and extract compact sources against 
a strong and variable background. In this way we obtained five lists of sources, corresponding 
to the five bands of the Hi-GAL survey, that have been merged adopting the association radii 
reported in Table~\ref{tab_bandcomplementing} for the {\it Herschel} photometry. In assembling 
this catalog the merging process proceeded from long to short wavelengths, updating the 
position of each merged source with the coordinates corresponding to the source actually 
associated at the shortest wavelength, with the aim to retain the position associated 
to the shortest detected wavelength. 

Subsequently, to evaluate accurate bolometric quantities, we also extended the spectral 
information on these FIR sources by complementing the merged {\it Herschel} catalog with {\it Spitzer} 
and WISE fluxes, at shorter, and with 1.2 mm SIMBA fluxes \citep{Mas07}, at longer wavelengths. 
In doing this we always adopt the positional requirements reported in Table~\ref{tab_bandcomplementing} 
to associate complementary fluxes to the fiducial positions in the multiband catalog of the 
{\it Herschel} sources.

As for the previous MIR catalog, the merging procedure allows the possibility of multiple source 
associations, within the constraints in Table~\ref{tab_bandcomplementing}), a problem that will 
be taken into account in the subsequent analysis of this FIR catalog.

%%%%%%%%%%%%%%%%%%%%%%%%%%%%%%%%%%%%%
\section{Sources Selection}    \label{Sources_selection}
To obtain a census as complete as possible of the protostellar content in the VMR-D cloud 
we exploit both the MIR and FIR catalogs based on the {\it Spitzer} and {\it Herschel} 
observations, respectively. Because in the following analysis we select candidate 
protostars from these catalogs, here we separately illustrate the selection procedure 
and the classification criteria adopted to characterize the selected sources.

\subsection{MIR sources}   
A preliminary analysis, limited to the {\it Spitzer} sources detected in all the four 
IRAC bands, was already presented in Paper I. Here we enlarge the data base to 
all the sources detected with photometric uncertainties $\sigma<0.2$ mag in at least 
three out of the four IRAC bands, augmented with those sources that, detected only in two IRAC 
bands, are also detected in the MIPS 24~$\mu$m band. As a final addition, with the aim to consider 
potentially interesting objects, we also include a few special cases that, detected in only one 
IRAC band, are particularly bright at $\lambda=24~\mu$m and show a spectral slope that is compatible 
with a Class I SED.

While including sources detected in only three wavelengths helps mitigate possible selection 
effects due to different sensitivities in the IRAC bands, considering also objects detected in only 
two IRAC bands and in the MIPS 24~$\mu$m  band allows us to include further sources that, for 
different reasons, could appear faint in particular wavelengths as, e.g., transition disks 
with large inner holes or unusual circumstellar geometries that could escape detection 
in some IRAC bands. The last addition of sources detected only in one IRAC band 
aims to include potentially interesting sources that could represent deeply embedded 
objects barely visible in the IRAC bands or even Class II sources with inner disk cleared from dust. 

In this way we extracted a total number of 16391 sources whose distribution in the different 
typologies is shown in the first row of Table~\ref{tab_sourceselection}. These constitute the 
working data base in which we search for candidate YSOs by using specific selection criteria aimed to 
exclude both Galactic and extragalactic contaminants. 

\subsubsection{Selection phase}  % Section: source selection
The first obvious step in selecting bona fide YSOs is to exclude from further investigation all 
those sources that could be interpreted as reddened stellar photospheres. To this aim, 
following \citet{Har06}, we compared the observed with the model fluxes, based on the 
stellar photosphere Kurucz-Lejeune models, taken from the SSC's Star-Pet tool 
\footnote{http://ssc.spitzer.caltech.edu/warmmission/propkit/pet/starpet/index.html}.
These fluxes are given just as they would be observed in the IRAC and MIPS bands and are 
then directly comparable with our observational data. 
In this respect, we scrutinized all the sources with at least three detections in the 
$2.2\leq \lambda \leq 24~\mu$m spectral region and, considering the effects of the extinction 
law given by \citet{Fla07}, we found 12217 sources that can be reasonably fitted with a reduced 
$\chi^2 < 2.5$ and then represent, most probably, reddened photospheres that have been consequently 
excluded from further consideration. With this choice the probability that some reddened normal 
star still contaminates our sample despite this filtering is less than $\sim 10\%$, a risk that 
will be further mitigated by the subsequent selection steps. 
In Figure~\ref{fig_dereddening} we show the distribution of the parameters involved in this 
procedure, noting that the bulk of the sources excluded by the adopted cut in $\chi^2$ 
involves essentially objects with slopes $-3.5~\lesssim~\alpha~\lesssim-1.5$ (bottom right panel) .  

After this preliminary step (Step 1 in Table~\ref{tab_sourceselection}) we are left with 4174 sources 
that have been further filtered adopting the selection criteria used by \citet[][see their Tab.1]{Har07} 
to identify extragalactic contaminants on the basis of the 4.5, 8.0, and 24~$\mu$m photometry. 
In this phase we also discarded sources with both color [4.5]-[8.0]~$<$~0.7 and spectral slope $\alpha~<-1.6$ as 
representing normal background stars remaining even after the previous dereddening phase and visible 
in Figure~\ref{fig_Har07} as a cloud of small black points. Applying these criteria to our sample 
we eliminate 2217 sources whose different typologies are reported in the row labeled ``Step~2'' 
in Table~\ref{tab_sourceselection}.

Excluding also these objects we count 1957 sources that have been further scrutinized with 
additional color-color and color-magnitude criteria similar to those devised by 
\citet[][their appendix A]{Gut09} to identify, and then exclude, further extragalactic as 
well as Galactic contaminants. Following these authors our sources have been examined in three 
different phases, the first consisting in the identification of sources whose colors are compatible 
with PAH contamined galactic and extragalactic sources, AGN, and protostellar shocks. In this phase 
484 sources (Step~3 row in Tab.~\ref{tab_sourceselection}) were flagged as contaminants so that, 
excluding them from further consideration, we remain with 1473 sources. 
In Figure~\ref{fig_Gut_shock_pah} the [3.6]-[4.5] vs [4.5]-[5.8] diagram is shown with 
the two regions, delimited by the continuous line, corresponding to the colors appropriate for 
shocks (upper left) and PAH contamined (lower right) sources \citep[see also][ fig.15]{Gut09}. 

As in the second phase of \citet[][ appendix A]{Gut09}, to include possible YSOs detected 
only in the [3.6] and [4.5] bands but not at 24~$\mu$m, we reconsidered the whole catalog to 
select all the cases with fluxes in JHK[3.6][4.5] bands, but requiring a photometric 
uncertainty $\sigma < 0.1$ mag in the JHK bands to minimize chances to pick-up 
sources only because of the additional uncertainty implied by the dereddening procedure at these 
shorter wavelengths. In this way we selected 48 sources that, 
%% se uso sigma<0.2 trovo 68 sorgenti ma nessuna passa i criteri. Da una verifica vedo che tutte 
%% queste sorgenti sono debolissime al limite di sensibilita' (1e-5 Jy) in banda 3.6 ma buone in 
%% banda 4.5.
however, do not satisfy the phase~2 criteria of \citet[][ appendix A]{Gut09} and then leave unaffected the 
total number of candidate YSOs.

In the third phase we recovered sources that, although detected in only one IRAC band, are however 
bright at 24~$\mu$m. These are potentially interesting sources because they could represent 
deeply embedded objects barely visible in the IRAC bands or even Class II sources with inner 
disk cleared from dust. Because of this we include in our sample of candidate YSOs all the 
sources with bright MIPS 24~$\mu$m photometry ([24] $<$ 7) to mitigate extragalactic contamination, 
and color [X] - [24] greater than the corresponding color expected for a SED with slope $\alpha = -0.3$, 
namely [X] - [24] $>$ 5.43, 4.77, 4.10, 3.22 for the four bands, where [X] is the only available 
IRAC band photometry. 
In this way we recovered 4, 4, 0, and 1 sources with good photometry in the IRAC 3.6, 4.5, 5.8, 
and 8.0~$\mu$m bands, respectively, increasing to 1482 the total number of candidate YSOs. All of 
these nine additional sources also show good MIPS 24~$\mu$m photometry with $\sigma < 0.2$~mag.
Finally, we complete the selection by excluding 12 sources showing a particularly irregular SED 
that cannot be represented by a single slope (see Table~\ref{tab_sourceselection} and 
section~\ref{MIR_classification} ), suggesting they are the result of source confusion. 
The final list of candidate YSOs selected out of the {\it Spitzer}-PSC catalog is then constituted by 1470 sources.

\subsection{FIR sources}   % Section: source selection 
Another starting point to select candidate protostellar sources in the VMR-D cloud can be 
adopted by using the {\it Herschel} observations that, arranged in a FIR catalog of compact sources 
as described in Section \ref{FIR_catalog}, can be exploited to analyze the SEDs. 
In this respect our approach to select candidate protostars is quite different because at 
FIR wavelengths the observations are more sensitive to colder objects, complementing 
in this way our view of the starforming process. 

Basically, candidate sources are selected by adopting specific prescriptions \citep[see also][]{Gia12} 
aimed to exclude cases that, also due to confusion problems, appear as spectrally inconsistent. 
To this aim we first eliminate from our FIR-catalog the ambiguities due to the presence of 
multiple associations, retaining only the nearest one to the fiducial catalog position, namely 
that determined in the band merging phase (see section~\ref{FIR_catalog}). 
Then, to prepare the analysis of the SEDs, we select the sources detected in at least 
three out of the four {\it Herschel} bands at $\lambda \ge 160~\mu$m, not peaking at 500~$\mu$m, 
and without a dip 
% in the SED 
between three adjacent wavelengths. In addition, we also require that these sources 
appear as resolved at $\lambda_{\mathrm{ref}}=250~\mu$m, a reference wavelength we assume 
to correspond to an optically thin regime \citep{Eli13}. All of these prescriptions are 
in view of exploiting a simple modified blackbody model to fit the {\it Herschel} FIR 
fluxes in the subsequent classification phase. 

The SEDs selected in this way have been in fact fitted with a simple modified blackbody 
model that assumes an optically thin regime for wavelengths $\lambda > \lambda_0$ and an 
optical depth described by $\tau=(\nu/\nu_0)^\beta$, where $\beta=2$ is appropriate for simple 
dust models and $\nu_0=c/\lambda_0$ is the frequency corresponding to $\tau=1$. 
The model fluxes are computed from the relation: 
\begin{equation} \label{mod_bb_model}
 F_\nu = (1-\mathrm{e}^{-\tau_\nu})~B_\nu(T_d)~\Omega
\end{equation} 
where $T_d$ is the dust temperature and $\Omega$ is the solid angle subtended by the source 
to the observer. The fact $\Omega$ can be observationally constrained clarifies our previous 
requirement for resolved sources so that, to this aim, we consider as spatially resolved 
all the sources with a deconvolved size satisfying: 
\begin{equation}
\mathrm{FWHM}_{\lambda,\mathrm{dec}}=\sqrt{\mathrm{FWHM}_\lambda^2-\mathrm{HPBW}_\lambda^2} ~>~ 0.6 \times \mathrm{HPBW}_\lambda 
\end{equation}
where FWHM$_\lambda$ is the observed source size and HPBW$_\lambda$ is the instrumental 
half power beam width at the given wavelength. Whith these choices, and assuming the VMR-D distance 
of 700~pc \citep{Lis92}, the beam size and the minimum resolved size corresponds to $\sim$0.06~pc and 
$\sim$0.02~pc, respectively.

However, constraining $\Omega$ with the deconvolved size at $\lambda_{\mathrm{ref}}=250~\mu$m 
is not strictly appropriate at all wavelengths, due to the increase of the apparent size when 
colder and larger volumes becomes ``visible'' at longer wavelengths. For this reason, before 
applying the fitting procedure, we normalized the fluxes obtained at longer wavelengths to the 
same angular size by scaling them with respect to the ratio of the deconvolved sizes, namely 
FWHM$_{\lambda_\mathrm{ref},\mathrm{dec}}$/FWHM$_{\lambda, \mathrm{dec}}$. 
This treatment is justified by the expectation that the radial density law in the cores 
is of the kind $\rho(r)\propto r^{-2}$ implying for the mass that M(r)$\propto r$ \citep{Shu77}, 
a situation that in absence of a strong temperature gradient suggests the aforementioned 
flux scaling at optically thin wavelengths \citep{Mot01, Gia12}. 
 
Finally, by using a fitting procedure based on the Levenberg-Marquardt technique \citep{Mar09} 
we derive best fitting values for T$_d$, and $\lambda_0$, in this way also checking that the best 
fits always occur for $\lambda_0 < 100~\mu$m, a posteriori justifying our assumption of an 
optically thin regime at $\lambda=250\mu$m.

%%%%%%%%%%%%%%%%%%%%%%%%%%%%%%%%%%%%%%%%%%%
\section{Classification} \label{sources_classification}
Here we exploit the fluxes collected both in the MIR and in the FIR catalogs to classify 
the protostars and characterize the SF process in the VMR-D cloud. In the following 
we discuss two different schemes we use for classifying {\it Spitzer} and {\it Herschel} 
selected sources, respectively.

\subsection{MIR sources} \label{MIR_classification}
As described in the introduction, here we adopt the YSOs classification scheme based on the 
spectral slope as evaluated in the MIR spectral range, with the addition of a further class, 
the Class 0. The latter has been introduced to include objects in the earliest phases 
that are very faint or even undetected in the MIR, being more easily detected in the 
FIR/submillimetric range.
The physical basis for this scheme relies on the evolutionary scenario in which a central 
object is initially surrounded by an envelope of gas and dust whose spectral importance decreases 
with time in favour of an emerging hotter circumstellar disk. 
The global spectral appearance consequently changes so that an external observer would classify 
the YSO depending on the MIR spectral slope and then on the elapsed time from the end of the 
envelope dominated phase corresponding to the Class 0 stage. However, due to the intrinsic MIR 
faintness during the Class 0 stage, the least evolved protostars cannot be efficiently identified 
in this spectral region so that other complementary classification schemes have been proposed 
that are based on different observables as, e.g., the bolometric luminosity 
and temperature \citep[see][]{Che95,You05}, quantities we shall consider later. 

%To evaluate 
Here we limit ourselves to classify the candidate YSOs by considering the spectral slope 
in the interval $2 \lesssim \lambda(\mu\mathrm{m}) \lesssim 24$ 
%for our candidate YSOs, 
and to this aim we use all the available fluxes collected in the MIR catalog. 
Two distibutions are shown in Figure~\ref{hist_alpha_vela.eps}, corresponding to the 
slopes obtained for the YSOs located ON and OFF-cloud, 869 and 356 cases respectively. 
Hereinafter with ON-cloud sources we mean those projected within the  
yellow contour line shown in Figure~\ref{fig_ds9_sources+coverage}, corresponding to 
N(H$_2$)=6.5$\times 10^{21}$ cm$^{-2}$ in the column density map obtained by exploiting 
the Herschel/Hi-GAL observations with the same procedure adopted by \citet{Eli13}. 
After regridding the 160, 250 and 350~$\mu$m maps onto the 
pixels of the 500~$\mu$m map, we computed the corresponding cold dust temperature 
and column density maps. This has been done by fitting the modified black body model 
in Equation~(\ref{mod_bb_model}) to the intensities observed at the different wavelengths 
in each pixel of our maps, assuming the optically thin case as a justified approximation 
at these wavelengths. 
Note however that in the region outside the {\it Herchel}-PACS coverage 
only the three SPIRE fluxes are available and then here we derive 
less accurate column density values. 

The final contour line obtained, beyond delimiting the cloud that is visible on 
our maps, corresponds to a visual extinction of A$_V \sim 3.5$ that is 
approximately the extinction beyond which the largest number of protostars are 
found in the Orion~A molecular cloud \citep{Lad13} and the column density distribution 
in the Auriga I cloud enters the low density region of the cloud itself \citep{Fro10}. 

Conversely the OFF-cloud sources, that we consider as representative of the background,  
are those located outside the previous contour line. 
Our cloud is embedded in a larger complex so that what we 
define background here could not represents the true Galactic background, 
but instead should be considered a mixture of two components, one associated to 
the same Vela cloud and the other to the field. Despite this ambiguity the subtraction 
of the OFF from the ON-cloud component, weighting for the corresponding solid angles, 
remains appropriate to estimate the consistency of the ``genuine'' ON-cloud population.

The two histograms in Figure~\ref{hist_alpha_vela.eps} show that earlier classes 
are relatively more abundant in the ON-cloud region and simply reflect the statistics, 
obtained for the different classes at the end of the source classification phase, 
that is reported in Table~\ref{tab_classes}. 
In this respect it is worth noting that the counts in Table~\ref{tab_classes} 
do not change significantly by moving the boundary line of the ON-cloud region 
up to $\sim$15\% in column density.

In this analysis we also discarded 12 sources, reported in Table~\ref{tab_sourceselection} as 
the Step 4 of the selection pipeline, because their SED is probably affected by 
confusion problems. For these sources the linear fit in the $\log({\lambda F_\lambda})$ 
vs $\log({\lambda})$ diagram could not reasonably account for their peculiar SED that 
appears monotonically decreasing up to 8~$\mu$m, reminiscent of a Class III, and then 
suddenly rising in the 24~$\mu$m band. 

We note however that, despite our efforts in eliminating the background contamination, 
the number of Class III sources remains the most uncertain because the colors of the 
normal stars closely mimic those of this class. Consequently, the number of the Class III 
sources should be considered more as an upper limit than as an intrinsic characteristic 
of the cloud.

\subsection{FIR sources}  % Section: source classification
Our FIR catalog is based on the {\it Herschel} sources and thus it is expected to be 
particularly rich in cold objects. In principle such objects can be either early 
protostars or quiescent cores and then it is essential to devise some criterium to 
distinguish protostellar from starless cores. This task can be accomplished 
by considering that while starless cores should be characterized by a single cloud 
temperature, the protostellar ones are instead expected to show a temperature gradient 
just because they harbour a protostar. To this aim we use the modified blackbody model 
described by Equation~(\ref{mod_bb_model}) 
to fit the FIR spectral points, but excluding the 70~$\mu$m fluxes since at this short 
wavelength the emission more likely results from warmer material and is then useful to 
trace the early protostellar phases. In other words the observed 70~$\mu$m flux is taken as 
discriminating the protostellar/starless nature of the sources depending on 
its compatibility with the flux expected from the model that best-fits the SED at the 
longer wavelengts. Consequently, objects showing a 70~$\mu$m flux exceeding the model 
flux by more than 3$\sigma$ are considered protostars while in all the other cases, including 
those without a 70~$\mu$m detection, we simply classify the sources as starless. 
In this phase, to allow a reasonable fit of the SED at $\lambda > 70~\mu$m, we 
only consider {\it Herschel} sources showing at least three detections in the 
$160 \le \lambda \le 500~\mu$m spectral range.

This classification procedure is illustrated in Figure~\ref{fig_herschel_fits} where two 
typical fits obtained with the model in Equation~(\ref{mod_bb_model}) are shown. 
In both cases we obtain a reasonable fit at $\lambda>160~\mu$m but in the right panel 
the 70~$\mu$m flux is clearly in excess with respect to that expected on the basis of the model. 
Interpreting this as a signature for the presence of a hotter embedded source, we classify this object 
as a candidate protostar. The converse is valid for the left panel showing a source we classify 
as starless.

\begin{figure}
 \epsscale{1.0}
 \plotone{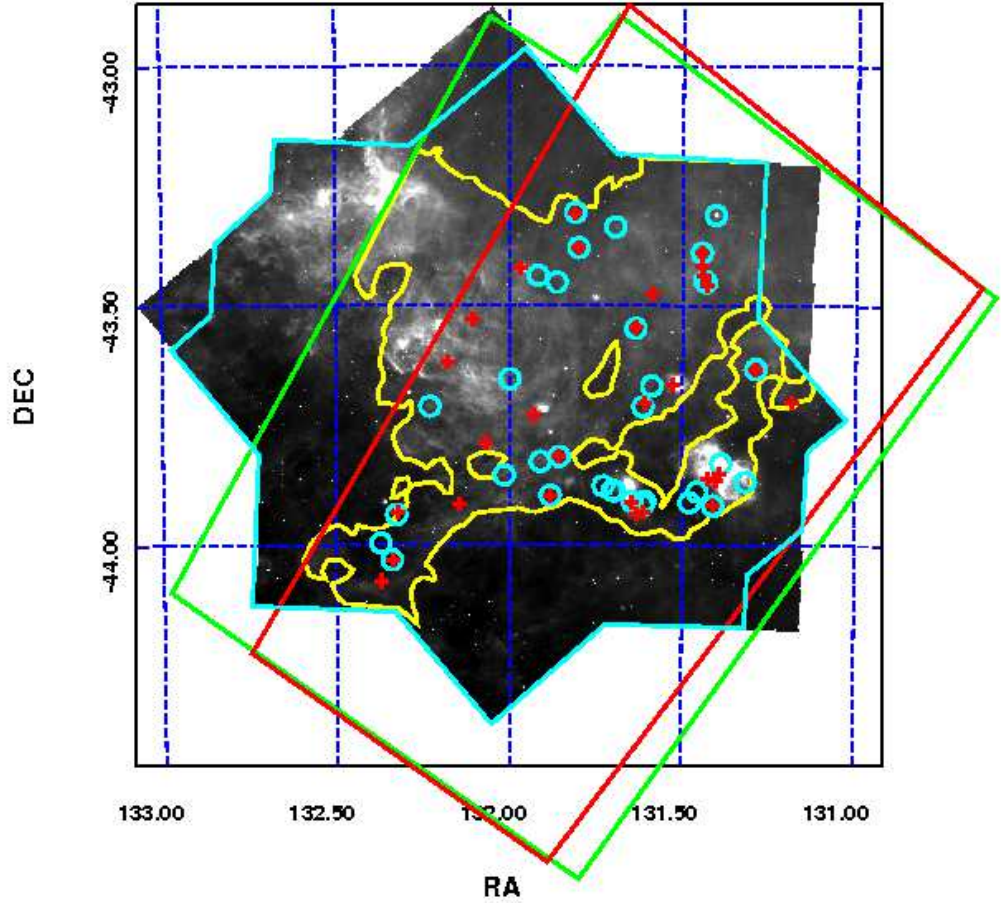}
 \caption{ IRAC 8.0 $\mu$m image of the VMR-D investigated field with the positions of the {\it Spitzer} 
           (cyan circles) and {\it Herschel} (red crosses) protostellar sources located ON-cloud and in 
           the common area surveyed by both {\it Spitzer} and {\it Herschel}. 
           The cyan polygon delimit the coverage of the {\it Spitzer}-PSC catalog (Paper I), 
           while the large tilted trapezoids delimit the coverage of the {\it Herschel} SPIRE (green) 
           and PACS (red) Hi-GAL observations analyzed in this work. 
           The yellow contour line corresponds to a column density N(H$_2)= 6.5 \times 10^{21}$ cm$^{-2}$
           and delimits the region we consider ON-cloud (see text). For a normal interstellar extinction 
           curve this corresponds to $A_V \sim 3.5$ mag.
           \label{fig_ds9_sources+coverage}}
 \end{figure}

\begin{figure}
 \epsscale{0.7}
 \plotone{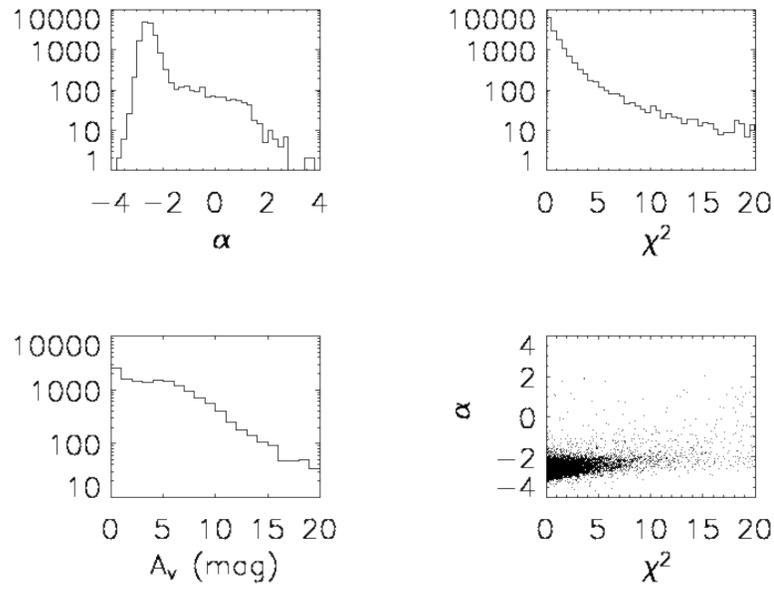}
 \caption{Best fit parameters obtained after comparing the observed MIR SEDs with the reddened photospheric 
          emission models (see text). Upper panels: distribution of the spectral slopes $\alpha$ (left) 
          and the reduced $\chi^2$ values (right) obtained by fitting the SEDs. Lower panels: the distribution 
          of the visual extinctions A$_V$ corresponding to the best-fits (left) and the $\alpha$ vs $\chi^2$ 
          scatter plot (right). \label{fig_dereddening}}
 \end{figure}

\begin{figure}
 \epsscale{0.6}
 \plotone{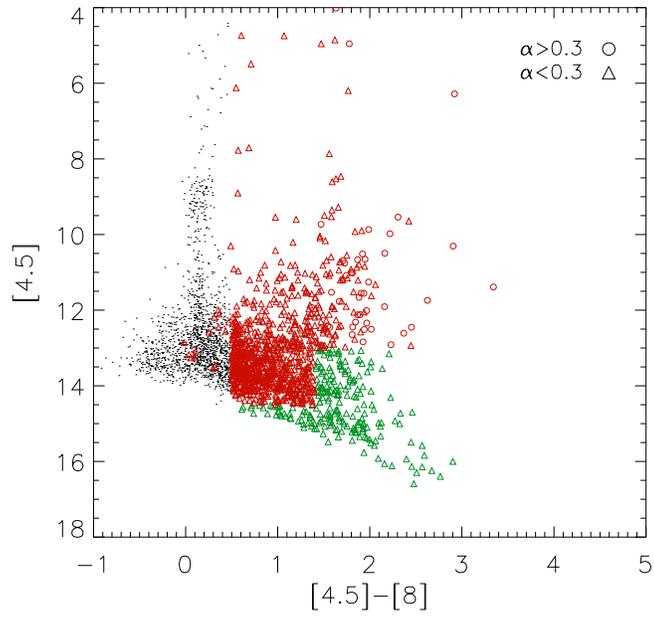}
 \caption{The IRAC color-magnitude diagram [4.5]  vs [4.5]-[5.8] showing the position of different objects. 
          Points represent background stars, triangles and circles are objects with $\alpha<0.3$ and $\alpha>0.3$, 
          respectively. Green and red symbols indicate objects that are rejected and passed, respectively, 
          after applying the \citet{Har07} criteria (see text). \label{fig_Har07}}
 \end{figure}

\begin{figure}
 \epsscale{0.6}
 \plotone{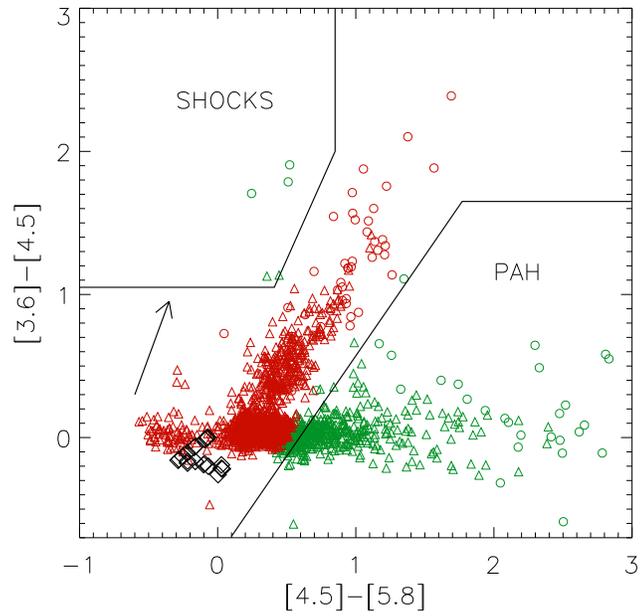}
 \caption{Color-color diagram exploiting IRAC fluxes to select candidate YSOs according to the \citet{Gut09} criteria. 
          The reddening vector is also shown, corresponding to a visual extinction A$_V$=50 mag and to 
          the reddening law determined by \citet{Fla07}. Symbols are as in Fig.\ref{fig_Har07}, except the 
          black diamonds, in the lower left corner, that represent the locus of the stellar photospheres. \label{fig_Gut_shock_pah}}
 \end{figure}

\begin{figure}
 \epsscale{0.7}
 \plotone{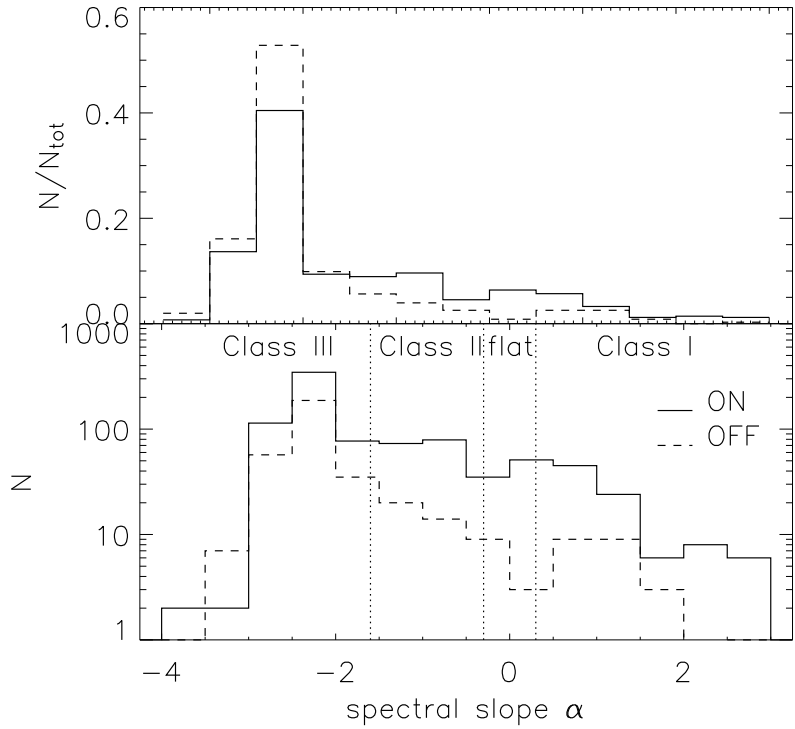}
 \caption{ Lower panel: the distribution of the MIR spectral slope $\alpha$ for the {\it Spitzer} selected candidate YSOs. 
           Sources located ON (solid line) and OFF-cloud (dashed line) are shown separately. The distinction is 
           referred to the position of the sources with respect to the contour line corresponding to 
           6.5$\times 10^{21}$ cm$^{-2}$ in the column density map obtained by the Herschel 
           observations (see text and Figure~\ref{fig_ds9_sources+coverage}). In the upper panel 
           the same distributions are shown normalized. \label{hist_alpha_vela.eps}}
 \end{figure}

\begin{figure}
 \epsscale{1.0}
 \plotone{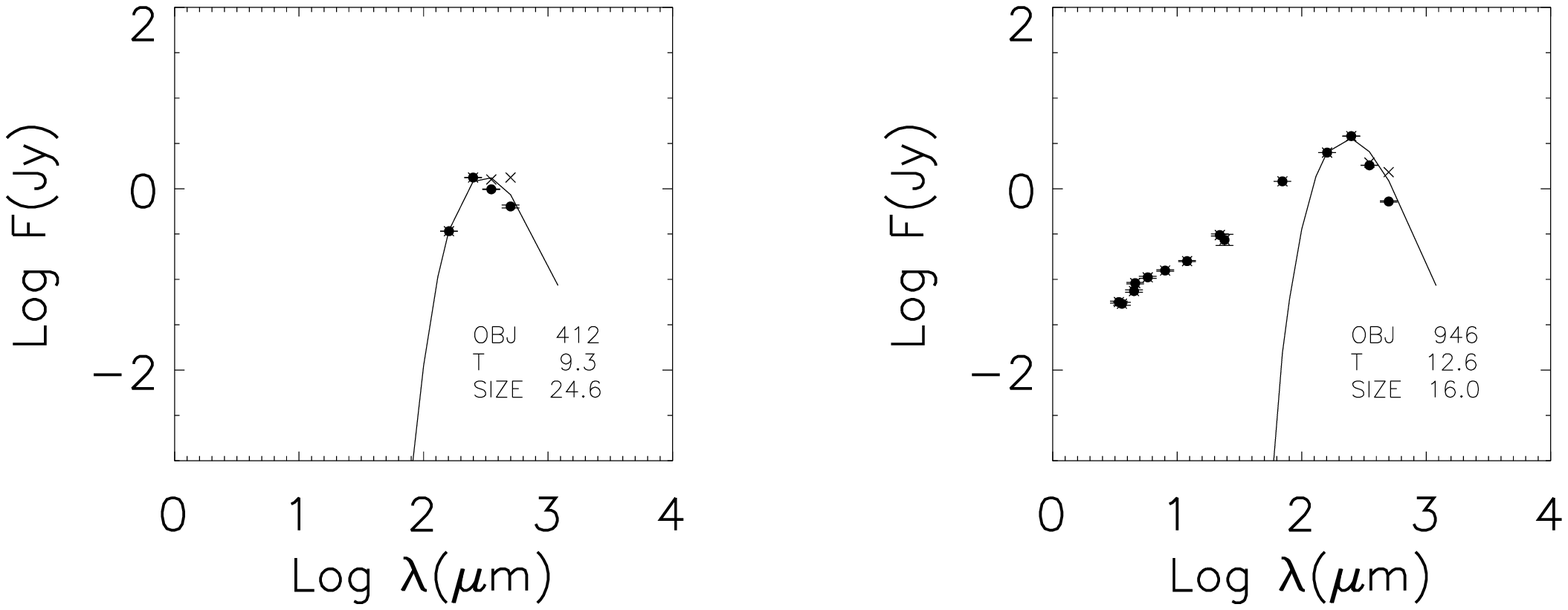}
 \caption{ Examples of SED fits obtained with a modified black body model for a typical starless 
           (left) and protostellar (right) source. The deconvolved size at 250 $\mu$m (in arcsec) 
           and the best fit temperature are also indicated.      \label{fig_herschel_fits}}
 \end{figure}

\begin{figure}
 \epsscale{1.0}
 \plotone{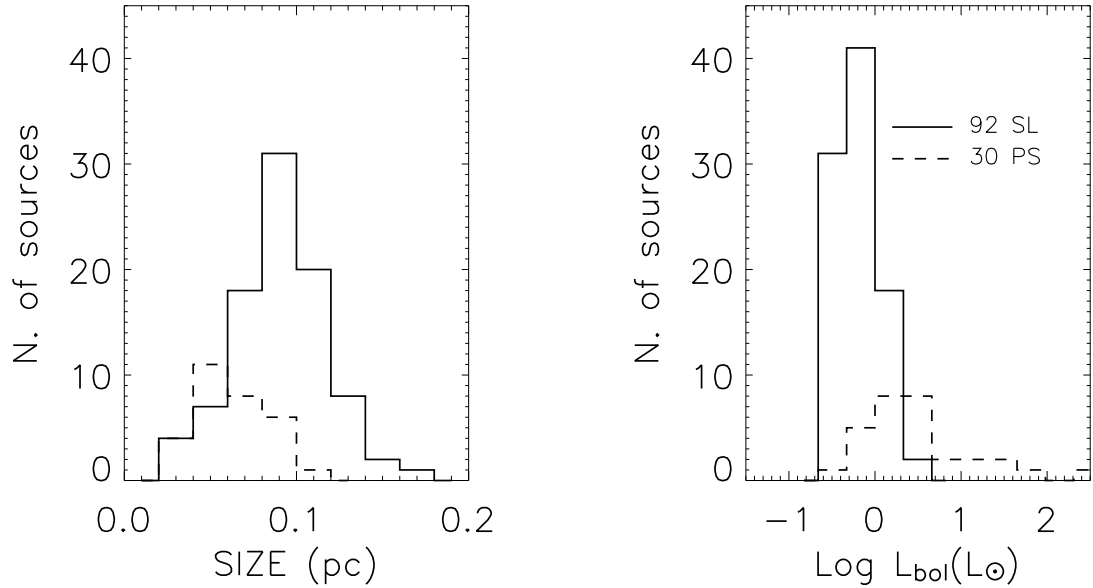}
 \caption{ Distribution in size (left) and luminosity (right) obtained for protostellar (dashed) 
           and starless (continuous) sources detected ON-cloud by {\it Herschel}. 
         \label{fig_size_lbol_hist}}
 \end{figure}

In this way we identified 31 protostellar and 100 starless sources in the overlap region covered 
by both {\it Herschel} and {\it Spitzer} observations (see Figure~\ref{fig_ds9_sources+coverage}) 
that reduce to 30 and 92 objects, respectively, when we consider only those located in the 
ON-cloud region. Their positions are shown in Figure~\ref{fig_ds9_sources+coverage} 
as red crosses, superimposed to the {\it Spitzer} 8~$\mu$m map of the VMR-D region, 
and their distribution in sizes is shown in the left panel of Figure~\ref{fig_size_lbol_hist}. 
The latter values are derived by using the deconvolved angular sizes observed at 250~$\mu$m and 
assuming a distance of 700~pc. 
Note that, because we need to consider the 70 $\mu$m 
flux to discriminate protostars, all these objects are confined to the region also covered 
by Herschel-PACS observations. 
In Table~\ref{tab_classes_herschel} we summarize the result of this classification by 
exploiting the bolometric temperature, introduced in the next section, as an alternative 
parameter to subdivide the {\it Herschel} protostars in classes.

%%%%%%%%%%%%%%%%%%%%%%%%%%%%%%%%%%%%%%%%%%%%%%%
\section{Bolometric quantities} \label{Bolometric_quantities}
Motivated by the difficulties in classifying protostars relying on either the 
MIR \citep{Lad87} or the submillimeter \citep{And93} spectral range, \citet{Che95} introduced 
a model-independent classification scheme based on the bolometric luminosity and the 
so-called bolometric temperature, originally defined by \citet{Mye93}. 
Because these two quantities can be derived observationally by using all the available 
fluxes, the resulting classification scheme is independent of the spectral characteristics 
in a predefined spectral range so that all the evolutive phases, including Class 0, 
can be more naturally included. The drawback is that bolometric quantities are more 
difficult to estimate because their accuracy depends on the observational 
coverage of the SED, particularly in the spectral region where most of 
the object's luminosity is emitted. 

In this respect the SEDs of the protostellar sources, especially for the earliest objects, 
typically peak at wavelengths longer than 24 $\mu$m so that the number of cases in which 
we can directly estimate the bolometric luminosity is usually limited by the availability 
of the FIR fluxes. In our case, because the VMR-D observations cover a wide spectral range, 
we can accurately compute the bolometric luminosity at least for all those sources that have 
been detected in a large spectral range.

\subsection{{\it Spitzer} selected sources}  %% bolometric quantities
We compute the bolometric luminosities of the {\it Spitzer} ON-cloud sources by using 
the fluxes reported in the MIR catalog, recalling that this also includes the 
associated {\it Herschel} fluxes, as described in section~\ref{MIR_catalog}. 
Preliminarly we purge the MIR catalog from multiple associations with counterparts coming 
from another catalog that satisfy the adopted positional constraint with more than one 
{\it Spitzer}-PSC source. In such cases we leave only the nearest association to the 
{\it Spitzer}-PSC position, ensuring in this way that each associated flux enters only 
one of the SEDs we consider. 
Furthermore, to obtain accurate luminosity values, we limited our attention to consider 
only the sources satisfying all of these requirements: 
\begin{itemize}
\item[-] the SED is composed by at least six observed spectral points 
\item[-] at least two FIR fluxes ($\lambda \ge 24~\mu$m) are available 
\item[-] at least three fluxes are available in the MIR ($\lambda<24~\mu$m) spectral region. 
\end{itemize}

As a consequence, because long wavelength fluxes are needed to accurately obtain 
bolometric quantities, we practically focus our attention to the ON-cloud sources 
falling within the region surveyed by {\it Herschel} PACS+SPIRE as is shown in 
Figure~\ref{fig_ds9_sources+coverage}.
With these criteria we practically select 33 objects whose SEDs are Class I and FS and whose 
photometry is reported in Table~\ref{tab_spitzer_protostars}. For these we 
assume a distance of 700 pc \citep{Lis92} to obtain the luminosity L$_\mathrm{bol}$ that is 
computed by using a power law piecewice integration between data points, 
including a Rayleigh-Jeans tail extending toward the long wavelengths. 
For the same objects we also compute the bolometric temperature given by \citep{Mye93}:
\begin{equation} \label{eq_Tbol}
T_\mathrm{bol} = 1.25\times10^{-11}~\overline{\nu} 
\end{equation}
where $\overline{\nu}$ is the mean frequency weighted with respect to the flux, defined as:
\begin{equation}
\overline{\nu} = \frac{\int_0^\infty \nu F_\nu d\nu}{\int_0^\infty F_\nu d\nu}
\end{equation} 
%% n. minimo di flussi per valutare Lbol: 6 di cui almeno 3 per determinare Lmir
%% dist. assunta 700 pc 

\begin{figure}
 \epsscale{1.0}
 \plotone{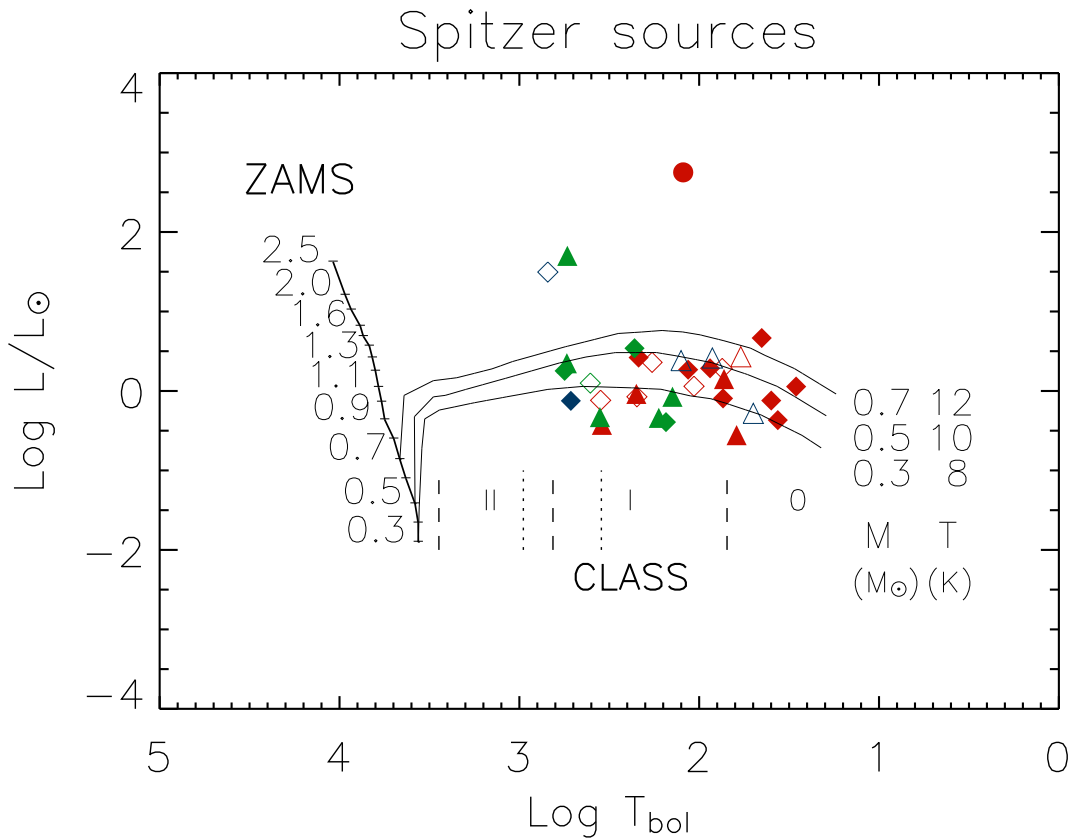}
 \caption{ Bolometric Luminosity-Temperature diagram obtained for the {\it Spizer} ON-cloud YSOs. 
           By using the classification criteria based on the MIR spectral slope red symbols correspond 
           to Class I, green is used for FS, and blue for Class II sources. 
           Different symbols denote different FIR spectral coverage: the filled circle is an object 
           with with {\it Herschel}-SPIRE fluxes as well as 24 $\mu$m and 1.2 mm fluxes. Diamonds 
           denote sources with 24 $\mu$m flux but undetected at 1.2 mm, with (filled) and 
           without (open) {\it Herschel}-SPIRE fluxes, respectively. Finally triangles are objects 
           without 1.2 mm flux and at least two fluxes in the 70--500 $\mu$m range, but with (filled) 
           and without (open) 24 $\mu$m flux, respectively.
%%%%%%%%%%%%%%%%%%%%%%%%%%%%%%%%%%%%%%%%%%%%5
%     Let(Sim)      | 24/22  70    250    350    500    1.2 | Ncasi| Comment  
%     A  (diamond)  |  Y     Y/N    N      N      N      Y  |   3  |
%     AB (square)   |  Y     Y/N    Y      Y      Y      Y  |   4  | 42629, 126339-> W22
%     B  (triangle) |  Y     Y/N    Y      Y      Y      N  |  11  | 115202 -> W22
%     NO (plus)     |  Y     Y      N      N      N      N  |  45  | 21825 ha 70+160 => Lbol
%     alt(cross)    |  Y     N    <almeno 1 detection>   N  |  12  |
%     N24(star)     |  N     <  almeno 2 flu > 24um   >  N  |   9  | solo 3 con Lbol calcolabile
%%%%%%%%%%%%%%%%%%%%%%%%%%%%%%%%%%%%%%%%%%%%%
           Evolutive tracks are taken from \citet{Mye98} and are labeled with the final stellar mass 
           and the initial temperature of the outer envelope. 
           In the bottom, with dashed lines, the correspondence between T$_\mathrm{bol}$ and spectral class 
           as defined by \citet{Che95} is reported for comparison. The boundaries of the FS sources 
           as suggetsed by \citet{Eva09} are also reported with dotted lines. 
           \label{Lbol_Tbol_MIR_ONcloud}}
 \end{figure}

The result of this procedure is presented in Figure~\ref{Lbol_Tbol_MIR_ONcloud} where the 
bolometric luminosities and temperatures, computed for the 33 ON-cloud sources 
satisfying the preceding criteria, are shown superposed on some evolutionary tracks 
taken from \citet{Mye98}. In the bottom of this figure we also report the correspondence 
between T$_\mathrm{bol}$ and the evolutive classes suggested by \citet{Che95} and \citet{Eva09}. 
It is noteworthy that the VMR-D cloud shows a relatively large number of sources 
(7 out of 33) that, classified as Class I (6 sources) and FS (1 source) according 
to their spectral slope in the MIR spectral range, show however T$_\mathrm{bol}<70$~K 
suggesting instead that they could well be Class 0 objects as far as the bolometric 
temperature is concerned.
This result confirms that the {\it Spitzer} sensitivity is sufficient to 
detect at least some of the Class 0 objects \citep{Dun14} and in fact we find that in the 
VMR-D region $\sim 20$\% of the {\it Spitzer} selected protostars can well be Class 0 sources.

\subsection{{\it Herschel} selected sources}  %% bolometric quantities
{\it Herschel} observations are more sensitive to the early protostellar phases 
\citep[see, e.g.,][]{Stu13} and then give us the opportunity to complete our information 
on the early objects. 
In this perspective the bolometric temperature and luminosity for the 30 protostellar 
and the 92 starless {\it Herschel} ON-cloud sources have been computed by using the 
merged FIR catalog described in section~\ref{FIR_catalog}. However, this catalog also 
includes complementary fluxes in the MIR spectral range so that we always use all the 
available fluxes, that for the ON-cloud protostars are reported in Table~\ref{tab_herschel_protostars}, 
to obtain the source luminosities. The resulting luminosity distribution is shown in the right 
panel of Figure~\ref{fig_size_lbol_hist} where a trend is apparent for the protostars 
to be, on average, more luminous than the starless cores. As expected, an opposite trend 
can be seen in the size distribution shown in the left panel where the protostellar 
sources appear as more compact objects with respect to the starless ones. 

\begin{figure}
 \epsscale{1.0}
 \plotone{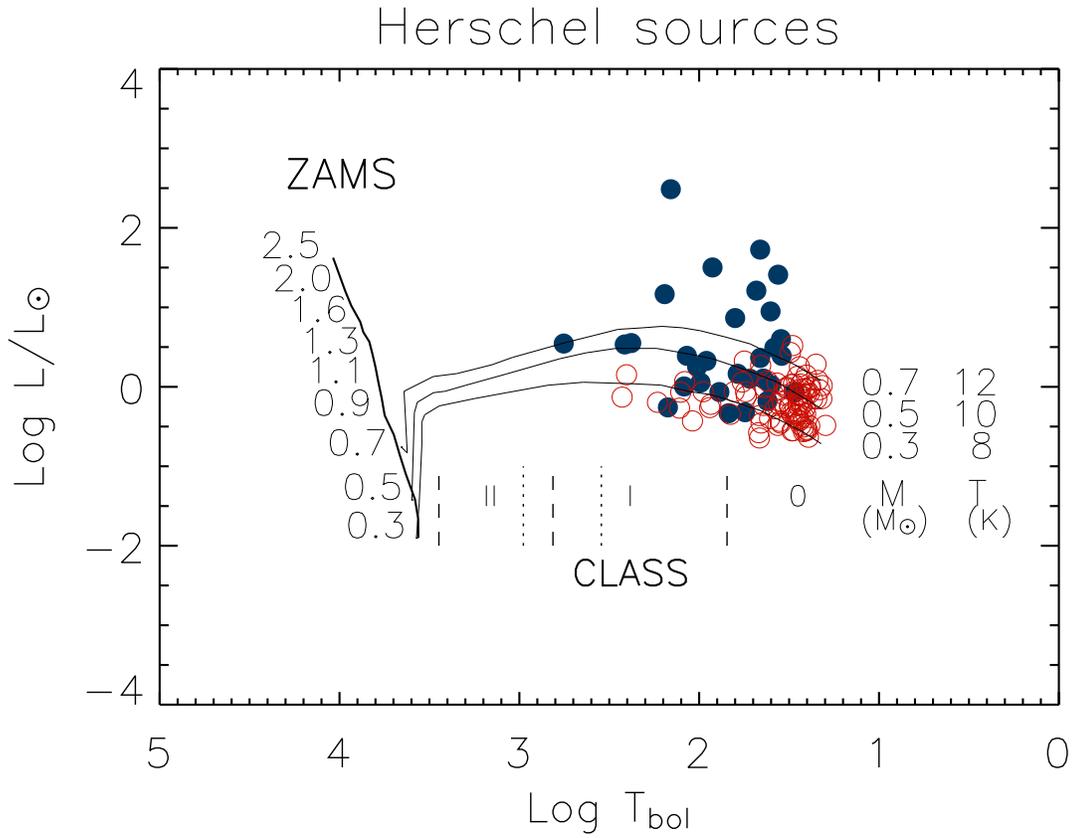}
 \caption{ Bolometric Luminosity-Temperature diagram obtained for sources detected by {\it Herschel}. 
           Filled blue circles are used for protostars, while open red circles represent 
           starless sources. The statistics of the protostars is summarized in 
           Table~\ref{tab_classes_herschel}, their fluxes in Table~\ref{tab_herschel_protostars}, 
           and their positions in the VMR-D cloud are shown in Figure~\ref{fig_ds9_sources+coverage}.
           Ten protostars, listed in Table~\ref{tab_common_protostars}, are in common with Fig.~\ref{Lbol_Tbol_MIR_ONcloud}. 
           \label{Lbol_Tbol_FIR_ONcloud}}
 \end{figure}

By computing the bolometric temperatures from Equation~(\ref{eq_Tbol}) we also obtain the
corresponding luminosity-temperature diagram that is shown in Figure~\ref{Lbol_Tbol_FIR_ONcloud}.  
It can be directly compared with Figure~\ref{Lbol_Tbol_MIR_ONcloud} because both 
refer to the same region of the VMR-D, namely the intersection between the 
{\it Herschel} coverage, the {\it Spitzer}-PSC catalog, and the ON-cloud region that is 
shown in Figure~\ref{fig_ds9_sources+coverage}. 
In the same figure the positions of both the {\it Spitzer} and {\it Herschel} selected 
sources are also shown.

Among these {\it Herschel} selected protostars we also find cases with T$_\mathrm{bol}~<~70$~K 
that can be classified as Class 0 objects, even though in this case they are relatively more 
abundant (17 out of 30 protostars) than in the {\it Spitzer} selected sample. 
This is clearly as expected because the earlier phases, corresponding to colder objects, 
are more easily detected at the longer wavelengths of the {\it Herschel} observations. 

It is however noteworthy that 10 out of the 30 {\it Herschel} protostars are also 
detected by {\it Spitzer} so that the total number of protostars in this region 
amounts to 53 sources. We then find that the {\it Herschel} observations are necessary 
to reveal 20 new protostars, corresponding to approximately one third of the total population 
in the surveyed part of the VMR-D cloud. 
 
For the protostars in common, in Table~\ref{tab_common_protostars} we report the classification 
determined by using the two schemes parameterized by the MIR spectral slope \citep{Lad87}
and by the bolometric temperature \citep{Che95, Eva09}, respectively. 

Considering now that the angular extension of our ON-cloud investigated region is 
$\sim 0.48$~deg$^2$ and that we adopted a nominal distance of 700~pc for the VMR-D, 
we find a protostellar surface density of $\sim 0.7$~pc$^{-2}$. 
A comparison with the star formation scaling law derived by \citet{Lom13} for the 
Orion molecular cloud suggests that the VMR-D protostellar density would be compatible 
with a visual extinction  $A_V \sim 5.5$, a value we find inside our ON-cloud 
region delimited by $A_V\sim 3.5$.

%%%%%%%%%%%%%%%%%%%%%%%%%%%%%%%%%%%%%%%%%%%%%%%%%%%%%%%
\section{Discussion} \label{Discussion}

A preliminary analysis of the young stellar population in the VMR-D cloud was presented in Paper I, 
but limited to the sources detected in all the four {\it Spitzer}-IRAC bands. 
Now, considering the full sample of the {\it Spitzer}-PSC sources and adopting a more demanding selection 
procedure, we obtain a more complete and reliable set of candidate YSOs to derive more accurate 
global properties of the SF in this cloud. 
Table \ref{tab_comparison} summarizes our results and, for the sake of comparison, also reports 
the results obtained by \citet{Eva09} for the SF regions of the c2d {\it Spitzer} legacy program. 
The comparison with Paper I shows that, despite the addition of sources detected in 
only three or even two IRAC bands, the number of FS and Class II candidate YSOs is 
similar as a result of adopting more stringent criteria to reject reddened photospheres 
in the selection pipeline (see Tab.~\ref{tab_sourceselection} and Fig.~\ref{fig_dereddening}). 
On the other hand, we select more Class I objects because a significant number 
of additional sources, detected in three or two IRAC bands, cannot be fitted by 
reddened photospheres and pass all the subsequent color-color and color-magnitude 
selection criteria discussed in Section~\ref{Sources_selection}. 
These are mainly sources detected at 3.6 and 4.5~$\mu$m, namely in the two more sensitive 
IRAC bands that are approximately ten times more sensitive than the 5.8 and 8.0~$\mu$m bands. 
The flux distribution of these additional sources peaks around 0.1 mJy showing that 
at these wavelengths they are faint objects near the completeness limit of the {\it Spitzer}-PSC 
catalog (see Tab.~\ref{tab_bandcomplementing}). Note however that, for the sake of a 
robust detection, these sources have been included in our analysis only if they also 
show a MIPS~24~$\mu$m counterpart. 

Similarly, the number of Class III objects is noticeably larger than that quoted in Paper I. 
In this case the additional sources are mainly detected only in the first three IRAC bands. 
Despite a large fraction of these three-band sources are actually discarded as reddened 
photospheres, a still consistent number survive our selection steps, showing fluxes brighter 
than 0.1 mJy with a distribution peaking at 0.4, 0.3, and 0.25 mJy in the first three 
IRAC bands, respectively. 
In this respect we must consider that the VMR-D line of sight crosses the Galactic plane and 
thus is particularly contamined by giant stars that, because of their thin dust envelopes, 
can closely mimic a Class III SED, making it difficult to distinguish them from genuine YSOs. 
In this respect we recall that also the 8~$\mu$m PAH emission, which can be produced 
in carbon rich red giant envelopes as well as in blob of nebulosity, could bias the 
selection of Class III sources in particular for the faint objects. 
Despite our efforts to mitigate this problem by adopting appropriate color constraints 
in Section~\ref{Sources_selection} and by subtracting the surface density of the 
YSOs seen outside the cloud (see Table~\ref{tab_classes}), we consider that the number of 
Class III sources quoted in Table~\ref{tab_classes} and \ref{tab_comparison} is the most uncertain. 

\subsection{Comparison with other SFRs}
\begin{figure}
 \epsscale{0.7}
 \plotone{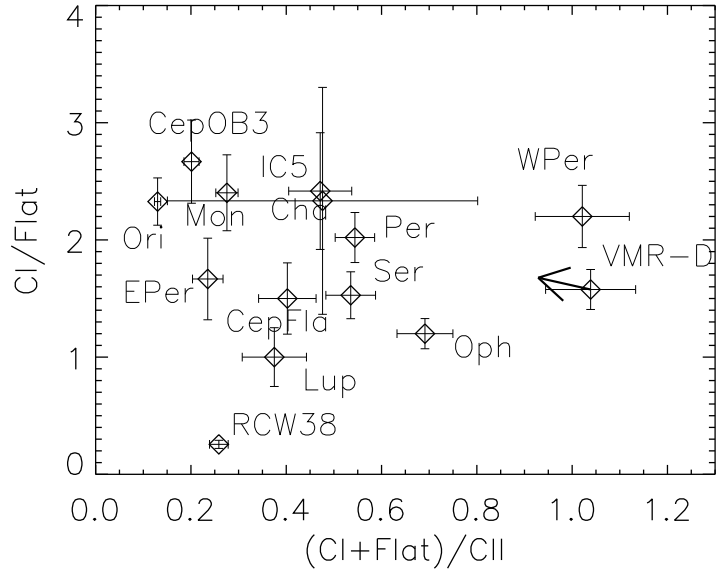}
 \caption{ The Class I/FS versus (Class I+FS)/Class II number ratio for Spitzer sources detected in 
           different clouds. VMR-D and WPer show a larger content of Class I and FS sources with 
           respect to the other SFRs. An arrow shows the displacement produced in the VMR-D 
           position by correcting for an extinction of A$_\mathrm{V}=5$ mag. 
           Data for other clouds are taken from \citet{Hsi13} for the c2d SFRs (Cha, Lup, Per, 
           Ser, and Oph, as well as EPer and WPer representing the eastern and western regions in Perseus). 
           For RCW 38, Cep, and IC 5146 we used the data in \citet{Win11,Kir09,Har08} while for 
           Ori, Mon, and CepOB3 those given by \citet{Kry12}, respectively. Note, however, that 
           \citet{Kry12} use more restrictive criteria in selecting FS sources and this probably 
           reflects in the upper left position of Ori, Mon, and CepOB3 in this diagram.
           \label{fig_CI_flat_CII}}
 \end{figure}

To compare the relative abundance of the different classes in VMR-D with those 
characterizing other SFRs, we show in Figure~\ref{fig_CI_flat_CII} the number ratios 
for Class I, FS, and Class II sources obtained in the different studies 
based on {\it Spitzer} observations, 
along with their uncertainties evaluated by assuming Poisson statistics. 
In this diagram the Class I/FS ratio for the VMR-D is well in the range of the other SFRs, 
while the ratio (ClassI+FS)/Class II appears clearly displaced toward the largest values. 
A possible sistematic shift in this diagram could be produced by the interstellar reddening 
that tends to steepen the spectral slope and then to produce a bias in the classification 
favouring the earlier classes. 
However, reasonable values of the interstellar extinction toward VMR-D do not significantly 
influence the position in this diagram as is shown by the arrow in Figure~\ref{fig_CI_flat_CII}
that represents the effect of an interstellar extinction A$_\mathrm{V}= 5$ mag. Given the values 
estimated for the insterstellar extinction toward VMR-D \citep[A$_\mathrm{V}\sim 2$mag, ][]{Cam99,Jos05} 
we find that the offset position of the VMR-D in this diagram cannot be explained by 
this effect and then more probably reflects an intrinsic difference in the relative populations. 
That this can be the case is also shown by the work of \citet{Hsi13} who noted that 
the relative number of the YSOs in Perseus significantly change if one considers the 
whole region or, separately, the two subregions extending eastward and westward of the 
R.A.=54$^\circ$.3. To illustrate this point in Figure~\ref{fig_CI_flat_CII} we also report 
these two subregions as EPer and WPer, in this way showing how they diverge from the position 
occupied by the Perseus considered as a whole. 

Intuitively we can guess that, in terms of a single SF burst, early objects are initially 
favoured so that both Class I/FS and (Class I+FS)/Class II number ratios should be 
relatively high. The absolute value and the time dependence expected for these ratios are 
clearly related to the specific modeling of the physical processes involved. Here we 
can only say that, if the position in this diagram has an evolutionary meaning, 
the VMR-D cloud appears much more similar to WPer and Ophiucus than to the other SFRs. 
However, if the SF occurs in multiple episodes or proceeds as a continuous process, 
Figure~\ref{fig_CI_flat_CII} cannot be easily interpreted and theoretical modeling is 
decisive to disentagle the different effects, a matter that is however beyond the scope 
of the present work. 

This simplified analysis suffers however at least two limits that should be considered 
in future work: the possibility that Class II reddened disks can appear as FS sources 
\citep{Dun14} and the presence, among the Class I objects, of cases that could well be 
Class 0 on the basis of their bolometric temperature. 
The first possibility tends to move  the individual clouds toward the upper-left 
part of Figure~\ref{fig_CI_flat_CII} as could be the case for Ori, Mon, and CepOB3 
whose FS sources have been selected by \citet{Kry12} with criteria that are 
more restrictive than those used for the other SFRs.
The second point on the possible presence of some Class 0 among the Class I 
objects should not change the significancy of the diagram as long as we simply 
reconsider that Class 0+I constitute a single counting bin. 
In this sense it is however noteworthy the larger relative content of protostars 
in VMR-D, WPer and Oph, suggesting for these regions a more recent SF event.

Once the YSOs are identified, further relevant quantities can be derived that characterize 
the SF in our cloud. These are reported in Table~\ref{tab_SF_rate} and are obtained for 
a distance of 700~pc and under some simplifying assumptions as a mean mass of 0.5~M$_\sun$, 
a binary fraction of 0.5, and an age of 2~Myr \citep{Eva09}, the latter being an estimate 
of the time taken to pass through the Class~II SED phase. It is interesting that, as for 
the position of the VMR-D in Figure~\ref{fig_CI_flat_CII}, we also find a greater similarity 
of the SF rate value to that of the Ophiucus cloud than to those of the other SFRs. 
% ricordo che c'e' un legame stabilito da Chen 1995 tra tempo dal collasso e Tbol: 
%    log t(yr) = 1.7 log Tbol + 0.65 pm 0.45     Tbol <1000 K 
%    log t(yr) = 2   log Tbol - 1.0  pm 0.54     Tbol >1000 K

\subsection{Protostellar luminosity function}
\begin{figure}
 \epsscale{1.}
 \plotone{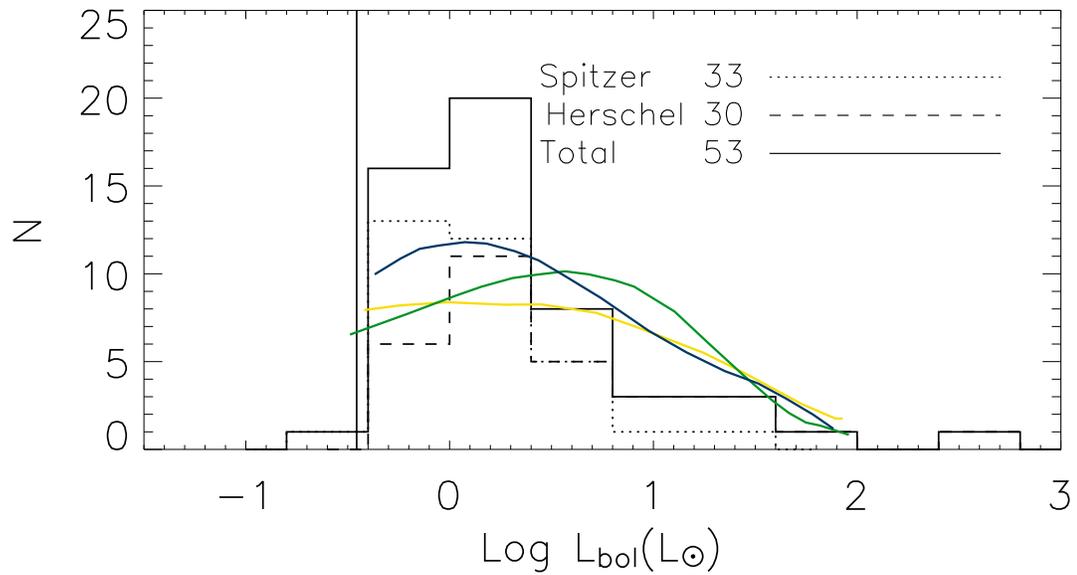}
 \caption{ PLF: {\it Spitzer} (dotted line) and {\it Herschel} (dashed line) 
           protostars located in the Spitzer/Herschel overlap region 
           of VMR-D (see Fig.~\ref{fig_ds9_sources+coverage}) and projected ON-cloud. 
           The total luminosity function is shown with a solid line accounting for 
           10 objects that are common to the two samples. The vertical line represents 
           the completeness limit based on the fluxes quoted in Table~\ref{tab_bandcomplementing}.
           Models from \citet{Off11} including competitive accretion (yellow), 
           turbulent core (green), and two component turbulent core model (blue), are also 
           shown for comparison. \label{fig_LF_combined}}
 \end{figure}

The PLF, determined by using all the protostars we find ON-cloud for which an 
accurate bolometric luminosity has been obtained in section~\ref{Bolometric_quantities}, 
is shown in Figure~\ref{fig_LF_combined}. 
% LAMBDA:       3.6     24.0      70    160   250   350   500 
%                                |---- 90% compl limit   ---- |
%               1.1e-5  6.3e-4   |0.21  0.67  1.05  1.32  1.95| Gia12 2012 photometry limits
%               1.1e-5  6.3e-4   |1.8   1.1   0.6   0.6   0.6 | Eli13 2013    "       limits
%
The completeness luminosity L$_\mathrm{com} \simeq 0.35$ L$_\sun$, shown as a vertical line, is 
evaluated on the basis of the completeness fluxes given in Table~\ref{tab_bandcomplementing}. 
In this figure we report only sources located ON-cloud in the overlap region observed by 
both {\it Spitzer} and {\it Herschel}, so that a comparison is possible between the ability 
of the two approaches to detect protostars. 
In particular we select 33 {\it Spitzer} and 30 {\it Herschel} protostellar 
sources respectively so that, considering that 10 sources are present in both samples, we 
conclude that in this region the {\it Herschel} observations contribute 20 additional 
protostars to those detected by {\it Spitzer}, leading up to 53 objects the total 
number of protostars.
Taken all together these produce a more complete and representative PLF that is shown in the 
same Figure~\ref{fig_LF_combined} with a continuous line. In the same figure three curves 
are also superimposed representing the PLF expected, after 1 Myr evolution, by models considering 
competitive accretion, turbulent core, and two component turbulent core, computed by \citet{Off11} 
with an upper limit for the final stellar mass of 3~M$_\sun$. 

A comparison with the PLF already known for other clouds suggests that the high luminosity 
tail extending beyond 100 L$_\sun$ makes the PLF in VMR-D more similar to those found 
in the high mass SFRs \citep{Kry12}. The most luminous source, namely 42629 in the 
{\it Spitzer}-PSC numbering, is a Class I in both classification schemes, based 
on the MIR spectral slope and the bolometric temperature, and is located in the center 
of a bright IR cluster \citep[IRS17,][]{Lis92, Mas00}. 
A more quantitative comparison with the luminosity distribution obtained in other SFRs is 
presented in Table~\ref{tab_statistics} that summarizes simple statistics corresponding 
to high and low-mass cases. 
The tabulated values suggest that the PLF in the VMR-D is more similar to that exhibited by 
the high mass SFRs, especially after correction of the reddening effect. 
Note however that our luminosities have not been corrected for this effect because they 
are essentially associated to Class 0 and I objects whose luminosity is dominated by the 
FIR contribution and is only marginally affected by the extinction. 

A much more important difference is due to the completeness luminosity that 
in our case is approximately ten times that quoted by \citet{Kry12}. These authors in 
fact, adopting a remarkable relationship they find between the bolometric and the MIR 
luminosity of the YSOs with $\alpha>0$, derive the bolometric luminosity for a large 
number of objects with a well sampled SED in the MIR, in this way exploiting the high 
sensitivity of {\it Spitzer}-MIPS 24~$\mu$m observations at the risk of an uncertain 
extrapolation. With this method they sensibly enlarge the number of sources useful to 
obtain a better coverage of the low luminosity tail of the distribution. 
In our case we adopt a more conservative approach and compute the bolometric luminosity 
of {\it Spitzer} selected sources only 
when at least two good quality FIR fluxes are actually observed in the 70--500 $\mu$m range, 
with the result of a larger completeness luminosity that is most probably the cause of the 
differences seen in Table~\ref{tab_statistics}.

\subsection{Envelope masses}
Because the PLF depends on both mass and time, its relationship to the initial 
mass function of the stars produced by the protostars is not straightforward. 
Ideally, one should know both the luminosity and the age of an object to infer 
its mass on the basis of some theoretical model. 
In principle, because the contributions to the observed luminosity come from a central 
object as well as from a surrounding disk and possibly an envelope, a detailed modeling 
of the global SED is to be used to disentangle the masses of the different components.

Nevetheless, in the early phases when the protostars appear as a Class 0/I, it is customary 
to simplify the problem of deriving the envelope mass by assuming that the FIR emission 
is optically thin and is then useful to trace the total envelope mass. 
In our case we could consider the subsample of the ON-cloud protostars with an observed 
500~$\mu$m flux, assuming that this is produced in an optically thin regime and adopting 
for the dust temperature the value obtained by fitting the {\it Herschel} fluxes at 
$\lambda\geq 160 \mu$m (see Section~\ref{Sources_selection}) with the model in 
Equation~(\ref{mod_bb_model}). However, given that most of our sources are resolved at 
{\it Herschel} wavelengths we can also obtain the masses following the derivation 
in \citet{Pez12} that involves the angular extension $\Omega$ of the emitting body:  
\begin{equation} 
M=\frac{D^2~\Omega}{k_\mathrm{ref}} \left( \frac{\lambda_0}{\lambda_\mathrm{ref}} \right)^2
\end{equation}

This expression allows us to derive the envelope mass for the 30 Herschel 
protostars (including the 7 cases without a 500 $\mu$m flux, see Table~\ref{tab_herschel_protostars}) 
because it does not depends explicitely on the observed flux at a given wavelength. 
This dependence is hidden in the fitting of the observed 
FIR fluxes with Equation~(\ref{mod_bb_model}) that gives also, as a by-product, 
the $\lambda_0$ value, namely the wavelength at which $\tau$=1. 
Similarly, we also extend this same approach to the starless cores 
detected in at least three {\it Herschel} wavelengths at $\lambda \ge 160~\mu$m.

\begin{figure}
 \epsscale{0.8}
 \plotone{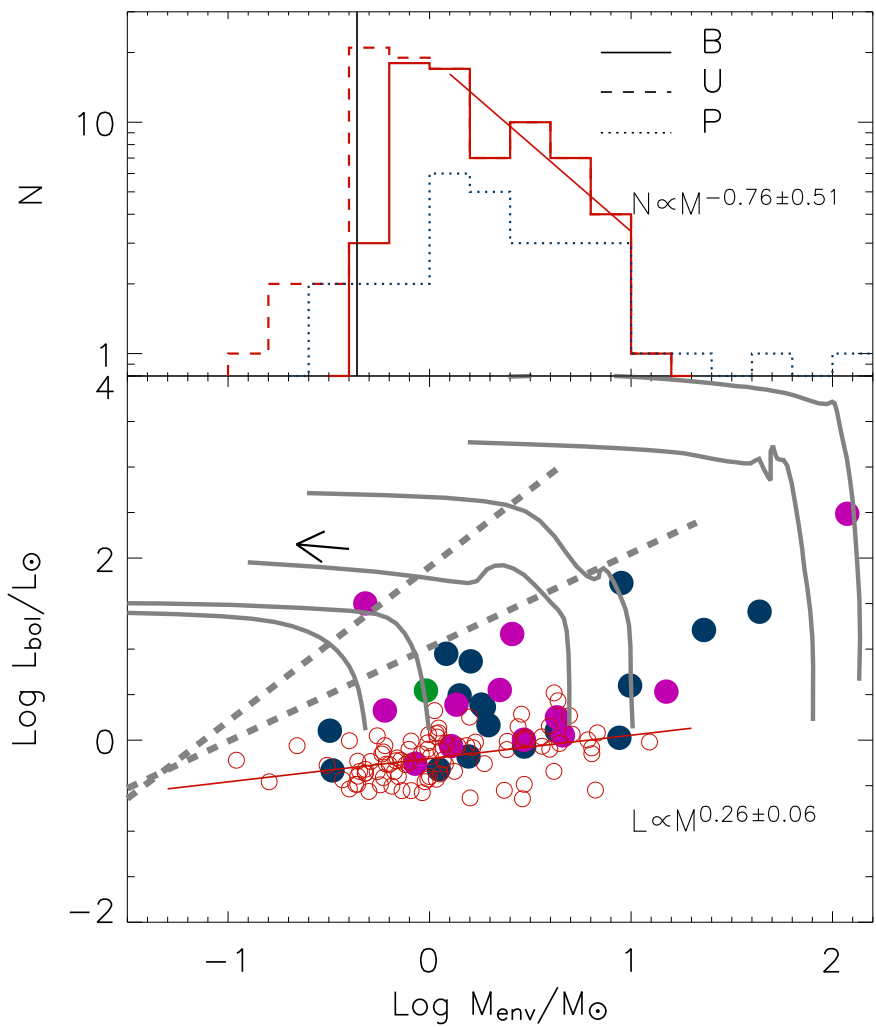}  % {fig_L_M_BAKY.eps}
 \caption{ Upper panel: mass distribution for starless and protostellar {\it Herschel} sources 
           detected ON-cloud. The red lines refer to the starless sources that are subdivided 
           in unbound (dashed) and bound (solid) cases. The dotted blue line indicates the 
           protostellar sources of Table~\ref{tab_herschel_protostars}, while the vertical 
           solid line is the completeness limit implied by the photometry limits in 
           Table~\ref{tab_bandcomplementing}. 
           Lower panel: L$_\mathrm{bol}$ vs M$_\mathrm{env}$ diagram: filled and open circles 
           represent protostars and starless cores, respectively.  
           Protostars classified as Class 0, I, and Flat Spectrum in Table 4, are indicated 
           with blue, magenta, and green filled circles, respectively.
           The superimposed grey lines represent the evolutionary tracks computed for 
           different initial envelope masses by \citet{Mol08} to follow the path of 
           a protostar in this diagram. The rising part corresponds to the main accretion 
           phase ending with the transition region between Class 0 and Class I objects.  
           This region is delimited by the two dashed lines taken from \citet{And00} that 
           ideally separate objects with M$_\mathrm{env} >$ M$_\star$ (lower) 
           from those with M$_\mathrm{env} <$ M$_\star$ (upper). The arrow shows the evolutionary 
           direction and the red straight line is a linear fit to the starless sources. 
           \label{fig_L_M_BAKY}}
 \end{figure}

The envelope masses have been then obtained adopting an emission coefficient 
$k_\mathrm{ref}$=0.1 cm$^{2}$ g$^{-1}$ of gas mass at $\lambda_\mathrm{ref}$=250~$\mu$m 
\citep{Hil83, Suu13} and a scaling law $k=k_\mathrm{ref} (\lambda_\mathrm{ref}/\lambda)^2$. 
In Figure~\ref{fig_L_M_BAKY} we present the resulting distribution of the envelope masses 
(upper panel) computed for both protostellar and starless sources. Because in the latter case 
the derived values represent the true total mass, it is interesting to compare the derived masses 
with the Bonnor-Ebert critical mass approximately given by $M_\mathrm{BE}\approx 2.4~R_\mathrm{BE}~a^2/G$, 
with $a$ being the sound speed at the source temperature. 
Such a comparison offers a simple way to separate prestellar from quiescent cores by 
simply assuming that masses M~$\gtrsim 0.5$~M$_\mathrm{BE}$ are bound and then 
should be considered as prestellar because they more probably evolve toward a protostar.

Assuming for $R_\mathrm{BE}$ the observed deconvolved radius at 250~$\mu$m we find that 
$M>0.5 M_\mathrm{BE}$ for 67 out of 92 ON-cloud sources that we then consider as bound 
and potentially prestellar. In Figure~\ref{fig_L_M_BAKY} the distribution of all the 
starless sources is shown as a dashed red line that merges, at $M>1M_\sun$, in the 
continuous red line corresponding to the prestellar component, suggesting that 
all the cores more massive that the solar mass are prestellar. Considering poissonian 
uncertainties and varying the binsize in reasonable limits (~10\% around a bin size=0.2) 
we obtain for masses $M>1M_\sun$ a slope of -0.76$\pm$0.51 where the uncertainty is 
dominated by the poor statistics. 
Even if this value is lower than that found in Vela-C molecular cloud by \citet{Gia12} 
and in the outer Galaxy by \citet{Eli13}, given the large uncertainty involved we 
cannot claim for meaningful differences. 

In the same figure we also show (bottom panel) the L$_\mathrm{bol}$ vs M$_\mathrm{env}$ 
diagram for the VMR-D protostars with superimposed the evolutive tracks computed by 
\citet{Mol08} for different initial envelope masses. For comparison we also report 
the starless sources noting that they populate the low luminosity part of the diagram 
and show a relatively smaller luminosity spread with respect to the protostars. 
A linear fit to their distribution in this diagram give a slope that is shallower 
than that found in both Vela-C \citep{Gia12} and the third Galactic quadrant \citep{Eli13}, 
suggesting possible differences in the clump/core forming mechanism.
On the other hand the protostars are more widely distributed in this diagram, 
the most of them being located on the rising branch of the evolutionary tracks 
corresponding to the main accretion phase. 
Following \citet{And00} we also report in this diagram two lines that delimit 
what is thought to be the transition zone separating the low luminosity branch in 
which M$_\mathrm{env} >$ M$_\star$, generally associated to the Class 0 phase, 
from the region in which M$_\mathrm{env} <$ M$_\star$ corresponding to the Class I 
and later phases. Remarkably, in most cases the protostars fall below the transition 
region suggesting a very recent star formation episode. However a similar distribution 
of the protostars is also seen in other SFR observed with {\it Herschel} 
\citep{Gia12,Eli13} and this seems to be unrealistic, raising a problem of consistency 
with the model. 
From an observational point of view, one possibility is that, due to our need to 
know at least three fluxes at $\lambda \ge 160~\mu$m to appropriately fit a modified blackbody, 
we introduce a selection effect favouring the earliest objects, a point that should 
be better considered in future work.

%%%%%%%%%%%%%%%%%%%%%%%%%%%%%%%%%%%
\section{Conclusions} \label{section_conclusions}
In this work we have studied the YSO population in the VMR-D cloud by exploiting both 
the {\it Spitzer}-PSC catalog, obtained from {\it Spitzer} observations and published 
in Paper I, and the recent {\it Herschel} observations carried out during the completion 
of the Hi-GAL key program involving part of this region. The main results can be summarized as follows: 
\begin{itemize}
\item[-] A list of 1470 candidate YSOs, identified out of the $\sim$170,000 {\it Spitzer}-PSC catalog sources, 
         has been obtained through a selection pipeline designed to exclude reddened normal stars, as well 
         as Galactic and extragalactic contaminants.
\item[-] YSOs have been subdivided in classes according to their spectral slope in the near-mid IR 
         and their relative numbers (86, 53, 136, and 282 for Class I, FS, II, and III, respectively), 
         obtained after subtraction of the estimated background population,
         have been compared with those characterizing other star forming clouds showing that the YSO 
         content in VMR-D is similar to that seen in Ophiucus and in the western part of Perseus.
\item[-] The {\it Spitzer}-PSC catalog has been complemented to obtain the largest data set on the continuum 
         spectrum of the sources, including fluxes from the 2MASS and WISE public surveys 
         as well as from new {\it Herschel} FIR observations and previous 1.2 mm SEST-SIMBA continuum 
         mapping of the VMR-D region.
\item[-] The {\it Herschel}/Hi-GAL maps ($\lambda=70$, 160, 250, 350,$\mathrm{and}~500~\mu$m) overlapping our 
         VMR-D region have been first used to identify the ON-cloud regions delimited by a 
         column density N(H$_2$)=6.5$\times 10^{21}$ cm$^{-2}$ and then analyzed to obtain a merged 
         catalog of compact FIR sources useful for searching additional protostars undetected by 
         {\it Spitzer}. This catalog, complemented with fluxes coming from the {\it Spitzer}-PSC 
         catalog as well as from WISE and the 1.2 mm catalog, has been exploited to find and 
         characterize 30 protostellar and 92 starless sources and to evaluate their bolometric 
         luminosities and temperatures. 
\item[-] Bolometric luminosities and temperatures have been also obtained for 33 {\it Spitzer} 
         selected ON-cloud protostars. These have been plotted in a $L_\mathrm{bol}$ vs $T_\mathrm{bol}$ 
         diagram showing that 7 (six Class I, and one FS) out of these 33 protostars 
         can actually represent Class 0 objects due to their low bolometric temperature (T$_\mathrm{bol}<$70 K).
\item[-] The distribution of the luminosities of both the {\it Spitzer} and {\it Herschel} selected protostars 
         show a bright outlier, corresponding to the same object {\it Spitzer}-PSC 42629 \citep[IRS17 in][]{Lis92}, 
         suggesting that VMR-D is also forming relatively high mass objects.
\item[-] The complete PLF of the ON-cloud region covered by both {\it Spitzer} and {\it Herschel} 
         observations has been obtained by merging together the 33 {\it Spitzer} and the 30 
         {\it Herschel} selected protostars. Given that 10 objects are in both samples we count 
         a total of 53 objects and find that {\it Spitzer} is able to detect approximately two-thirds 
         of the protostellar population actually present in VMR-D. This stresses the need for 
         using also the {\it Herschel} observations in studying the earlier protostellar phases. 
\item[-] The envelope masses have been obtained for all the {\it Herschel} protostellar sources 
         detected ON-cloud and for which an envelope temperature can be determined. A plot of the 
         masses versus the corresponding luminosities shows that, by comparison with evolutive tracks, 
         we are mainly sampling objects in the main accretion phase. A possibility is that a 
         selection effect is implied by our preference for sources also detected in the FIR, 
         a choice that is dictated by the need to obtain accurate bolometric quantities.
\end{itemize}




%\clearpage

%% TABLES:
\begin{deluxetable}{lcccccc}
\tabletypesize{\scriptsize}
%\rotate
\tablecaption{Catalogs considered. \label{tab_bandcomplementing}}
\tablewidth{0pt}
\tablehead{
\colhead{Name}                                & \colhead{2MASS} & \colhead{WISE}          & \colhead{{\it Spitzer}-IRAC+MIPS} & \colhead{Herschel}           & \colhead{SEST-SIMBA} \\
\colhead{Band}                                & \colhead{J H K} & \colhead{3.4 4.6 12 22} & \colhead{3.6 4.5 5.8 8.0 24 70}   & \colhead{70 160 250 350 500} & \colhead{1200} }
\startdata
Beam size (\arcsec)                            &    2.5 -- 3     &  6 --  12               &  1.7 -- 6                         & 7.6  12.3  18   25   36      &   24   \\ 
Association radius\tablenotemark{a} (\arcsec) & 0.6             &  0.9                    &  1                                & 4     8    11   16   21      &   12   \\
Completeness\tablenotemark{b} flux (mJy)                       & 0.76 1.23 1.67  & 4.9 2.7 3.2 5.3         &  0.05 0.05 0.29 0.4  2.0  500     & 1800 1100  600  600  600     &   20   \\  % old(2012) compl. lim.: 210  670  1050  1320 1950
Reference\tablenotemark{c}                    & 1               &  2                      &  3                                & 4                            &   5    \\
% Limiting flux (mJy @ 5$\sigma$)               & 0.38 0.33 .65   & 0.08 0.11 1 6           &  0.01 0.01 0.04 0.05  0.6 100     & 40   90    110  350  460 TBC &   20   \\
% Reference\tablenotemark{b}                    & 1               &  2                      &  3                                & 4, 5                         &   6    \\
\enddata
% 2MASS:  15.8, 15.1, 14.3 mag @ 10sig   ==> 0.76, 0.65, 1.3  mJy   >http://www.ipac.caltech.edu/2mass/releases/allsky/doc/sec2_2.html 
% WISE :  W1 W2 W3 W4          @ 5 sig   ==> 0.08, 0.11, 1, 6 mJy   >http://wise2.ipac.caltech.edu/docs/release/allsky/ (March 14, 2012) 
% *** Akari 9, 18 micron: 50, 90   (mJy) sono la detection limit a 5 sigma che corrisponde ad una completezza del 10% nel catalogo ***   
% ***        >vedi AKARI/IRC All-Sky Survey Point Source Catalogue Version 1.0-Release Note-(Rev.1) March 30 2010, pag 29 in basso    ***
% *** Akari 60, 90, 140, 160 micron: 1990, 320, 2000, 3800 (mJy) sono le detection limit a 5sigma "normal mode" per una completezza del 10% del catalogo ****
% ***        >vedi AKARI/FIS All-Sky Survey Bright Source Catalogue Version 1.0 Release Note March 30 2010, pag 23 
% BLAST  250 e 350 micron: 4 Jy ricavati da Fig.3 di: 2009-Nettefield_etal_ApJ_707_1824 ----------- 500 micron TBD.
% 
%% Text for table notes should follow after the \enddata but before 
%% the \end{deluxetable}. Make sure there is at least one \tablenotemark 
%% in the table for each \tablenotetext. 
% \tablecomments{Table \ref{tbl-1} ............... TBD ............}
\tablenotetext{a}{1-$\sigma$ positional uncertainty for 2MASS and WISE catalogs. For {\it Spitzer}, {\it Herschel} and SEST data 
the association radius is estimated from the references given.}
%\tablenotetext{b}{References: (1)\citeauthor{Skr06} \citeyear{Skr06}, (2)\citeauthor{Wri10} \citeyear{Wri10}, (3)Paper I, 
%               (4)\citeauthor{Pog10} \citeyear{Pog10}, (5)\citeauthor{Gri10} \citeyear{Gri10}, (6)\citeauthor{Mas07} \citeyear{Mas07}.}
\tablenotetext{b}{Completeness fluxes are meant for crowded regions: 2MASS 99\%, WISE 95\%, Spitzer 99\%, Herschel: 90\%, SEST: sensitivity limit.}
\tablenotetext{c}{References: (1)\citeauthor{Skr06} \citeyear{Skr06}, (2)\citeauthor{Cut13} \citeyear{Cut13}, (3)Paper I, 
               (4)\citeauthor{Eli13} \citeyear{Eli13}, (5)\citeauthor{Mas07} \citeyear{Mas07}.}
\end{deluxetable}

\begin{deluxetable}{lccccccccccc}
\tabletypesize{\scriptsize} 
%\rotate
\tablecaption{{\it Spitzer} Source counts through the selection pipeline. \label{tab_sourceselection}}
\tablewidth{0pt}
\tablehead{  \colhead{Step} & \colhead{1234\tablenotemark{a}} & \colhead{123X} & \colhead{12X4} & \colhead{1X34} & \colhead{X234}  &  \colhead{XX34}  &  \colhead{1XX4}  & \colhead{12XX} & \colhead{1X3X} & \colhead{X2X4}  & \colhead{Total}  } 
\startdata
0: first selection\tablenotemark{b} &    8852(436)   &  7210(28)      &     151(12)    &      19(12)    &       30(19)    &      9(9)     &         1(1)     &     116(116)  &       2(2)     &      1(1)       &  16391  \\
1: reddened stars                   &   -6110        & -6063          &      -9        &     -10        &      -15        &     -6        &         0        &      -3       &       0        &     -1          & -12217  \\
2: Har07 criteria\tablenotemark{c}  &   -2059        &     0          &    -142        &       0        &      -15        &      0        &        -1        &       0       &       0        &      0          &  -2217  \\
3: Gut09 criteria\tablenotemark{d}  &     -17        &  -467          &       0        &       0        &        0        &      0        &         0        &       0       &       0        &      0          &   -484  \\
4: undefined slope                  &      -9        &    -3          &       0        &       0        &        0        &      0        &         0        &       0       &       0        &      0          &    -12   \\ 
   Total                            &     657        &   677          &       0        &       9        &        0        &      3        &         0        &     113       &       2        &      0          &  1461\tablenotemark{e}   \\
\enddata
\tablenotetext{a}{Four digits are used to refer to the IRAC bands in which the sources have been detected: e.g. 12XX refers to  
                  sources detected in the first and second band only. The X23X column is not reported because no sources 
                  of this kind have been found.}
\tablenotetext{b}{Numbers in parenthesis are sources showing also MIPS-24~$\mu$m fluxes.}
\tablenotetext{c}{\citet{Har07}.}
\tablenotetext{d}{\citet{Gut09}.}
\tablenotetext{e}{Nine bright 24~$\mu$m sources, detected in only one IRAC band, have been subsequently added (see text).}
\end{deluxetable}

\begin{deluxetable}{lccccc}
%\tabletypesize{\scriptsize} 
%\rotate
\tablecaption{VMR-D Statistics of the {\it Spitzer} YSOs (MIR catalog). \label{tab_classes}}
\tablewidth{0pt}
\tablehead{ \colhead{Region} & \colhead{ON\tablenotemark{a}} & \colhead{OFF\tablenotemark{a}} & \colhead{OUT\tablenotemark{a}} & \colhead{TOT} & \colhead{ON-OFF\tablenotemark{b}} \\ 
~~~~~~~$\Omega$\tablenotemark{c}(deg$^2$) $\rightarrow$ & 0.54  & 0.52      & 0.14          & 1.20          &  0.54  } %     \\ 
 \startdata
             Class I        &   105(12\%)  &      22       &       22      &     149       &   82$\pm$ 11 (15\%)  \\
             Flat spectrum  &    58~(7\%)  &       6       &        9      &      73       &   52$\pm$  8 (10\%)  \\
             Class II       &   171(20\%)  &      40       &       33      &     244       &  129$\pm$ 14 (24\%)  \\ 
             Class III      &   535(61\%)  &     288       &      181      &    1004       &  236$\pm$ 28 (51\%)  \\ 
\enddata
\tablenotetext{a}{The columns ON and OFF refer to regions inside and outside the column density 
     contour line derived from the {\it Herschel} observations and shown in Figure~\ref{fig_ds9_sources+coverage}. 
     The OUT column refers to sources of the IRAC catalog outside the coverage of the {\it Herschel}-SPIRE 
     observations. }
\tablenotetext{b}{The difference has been weighted by the corresponding solid angles and the estimated uncertainty is poissonian.}
\tablenotetext{c}{Solid angle subtended by the different subregions.}
\end{deluxetable}
% ........................ 21 mar 2013
% per valutare la sensitivity cutoff in Lbol  (entro cui verosimilmente c'e' anche completezza) 
% usare l'idea di Kryukova pg 14..." 
% 1) assumere una SED slope steeper than 90% di tutte le sorgenti nella regione 
% 2) valutare Lmir di questa SED che finisce con una m24 pari a quella di cutoff (eq. 7)
% 3) valutare la Lbol con la eq. 8

\begin{deluxetable}{lcccccc}
\tablecaption{VMR-D Statistics of the {\it Herschel} protostars (FIR catalog). \label{tab_classes_herschel}}
\tablewidth{0pt}
\tablehead{ \colhead{Region\tablenotemark{a}} & \colhead{ON} & \colhead{OFF}    \\ 
~~~~~~~$\Omega$(deg$^2$) $\rightarrow$  & \colhead{0.48}    & \colhead{0.33}   \\
          } 
 \startdata
  Class 0\tablenotemark{b}  &   17    &  \nodata       \\
  Class I                   &   12    &    1           \\
  Flat Spectrum             &    1    &  \nodata   \\ 
\enddata
\tablenotetext{a}{ON and OFF have the same meaning as in Tab.~\ref{tab_classes} but the corresponding 
                  solid angles are further limited by the {\it Herschel}-PACS coverage (see text).}
\tablenotetext{b}{Based on the correspondence between bolometric temperature and classes (see text and 
                  Figure~\ref{Lbol_Tbol_FIR_ONcloud}).}
\end{deluxetable}

\begin{deluxetable}{lccccccccccccccccccccccccc}
%\rotate
\tabletypesize{\scriptsize}
\tablecaption{Photometry of the {\it Spitzer} selected protostars\tablenotemark{a} \label{tab_spitzer_protostars}}
\tablewidth{0pt}
\tablehead{
\colhead{ID}                        & \colhead{PSC\tablenotemark{b}}      & \colhead{RA}                         & \colhead{DEC}                       & \colhead{$F_{J}\pm\Delta F_{J}$}     & \colhead{$F_{H}\pm\Delta F_{H}$}     \\ 
\colhead{$F_{K}\pm\Delta F_{K}$}    & \colhead{$F_{W1}\pm\Delta F_{W1}$}  & \colhead{$F_{W2}\pm\Delta F_{W2}$}   & \colhead{$F_{W3}\pm\Delta F_{W3}$}  & \colhead{$F_{W4}\pm\Delta F_{W4}$}   & \colhead{$F_{3.6}\pm\Delta F_{3.6}$} \\ 
\colhead{$F_{4.5}\pm\Delta F_{4.5}$}& \colhead{$F_{5.8}\pm\Delta F_{5.8}$}& \colhead{$F_{8.0}\pm\Delta F_{8.0}$} & \colhead{$F_{24}\pm\Delta F_{24}$}  & \colhead{$F_{70}\pm\Delta F_{70}$}   & \colhead{$F_{160}\pm\Delta F_{160}$} \\ 
\colhead{$F_{250}\pm\Delta F_{250}$}& \colhead{$F_{350}\pm\Delta F_{350}$}& \colhead{$F_{500}\pm\Delta F_{500}$} & \colhead{$F_{1200}\pm\Delta F_{1200}$\tablenotemark{c}} \\
}
\startdata
%    NUM ID    RA         DEC         J           E_J         H           E_H         K           E_K         W1          E_W1        W2          E_W2        W3          E_W3        W4          E_W4        F36         E_F36       F45         E_F45       F58         E_F58       F80        E_F80       F24          E_F24       F70         E_F70       BLUE        E_BLUE      RED         E_RED       PSW         E_PSW       PMW         E_PMW       PLW         E_PLW       FMM_IRAC 
%    int int   real       real        real        real        real        real        real        real        real        real        real        real        real        real        real        real        real        real        real        real        real        real        real       real        real         real        real        real        real        real        real        real        real        real        real        real        real        real        real 
%              deg        deg         Jy          Jy          Jy          Jy          Jy          Jy          Jy          Jy          Jy          Jy          Jy          Jy          Jy          Jy          Jy          Jy          Jy          Jy          Jy          Jy          Jy          Jy          Jy          Jy          Jy          Jy          Jy          Jy          Jy          Jy          Jy          Jy          Jy          Jy          Jy          Jy          Jy   
    1                            &   8036                           & 131.28840                        & -43.63255                        &                       \nodata    &                       \nodata   \\
                       \nodata   &    0.00016    $\pm$   0.00005    &    0.00025    $\pm$   0.00009    &    0.0027     $\pm$   0.0003     &    0.0165    $\pm$   0.0013      &    0.000040   $\pm$   0.000004   \\
\enddata 
\tablenotetext{a}{Coordinates are given in degrees; fluxes ($F$) and their uncertainties ($\Delta F$) are given in Jy.}
\tablenotetext{b}{ID in the VMR-D PSC catalog (Paper I).}
\tablenotetext{c}{The uncertainty of the 1.2~mm flux is 20\% \citep{Mas07}.}
\end{deluxetable}

     \begin{deluxetable}{cccccccccccccccccccccccccc}
%\rotate
\tabletypesize{\scriptsize}
\tablecaption{Photometry of the {\it Herschel} selected protostars\tablenotemark{a} \label{tab_herschel_protostars}}
\tablewidth{0pt}
\tablehead{
\colhead{ID}                         & \colhead{RA}                         & \colhead{DEC}                        & \colhead{$F_{W1}\pm\Delta F_{W1}$}   & \colhead{$F_{W2}\pm\Delta F_{W2}$}   & \colhead{$F_{W3}\pm\Delta F_{W3}$} \\
\colhead{$F_{W4}\pm\Delta F_{W4}$}   & \colhead{$F_{3.6}\pm\Delta F_{3.6}$} & \colhead{$F_{4.5}\pm\Delta F_{4.5}$} & \colhead{$F_{5.8}\pm\Delta F_{5.8}$} & \colhead{$F_{8.0}\pm\Delta F_{8.0}$} & \colhead{$F_{24}\pm\Delta F_{24}$} \\
\colhead{$F_{BLUE}\pm\Delta F_{BLUE}$}& \colhead{$F_{RED}\pm\Delta F_{RED}$} & \colhead{$F_{PSW}\pm\Delta F_{PSW}$} & \colhead{$F_{PMW}\pm\Delta F_{PMW}$} & \colhead{$F_{PLW}\pm\Delta F_{PLW}$} & \colhead{$F_{1200}\pm\Delta F_{1200}$\tablenotemark{b}} 
}
\startdata
    1                             & 131.18478                         & -43.70255                         &   0.00221   $\pm$   0.00005       &   0.00141   $\pm$   0.00003       &   0.0025    $\pm$   0.0004     \\    
   0.046      $\pm$   0.001       &    0.00053    $\pm$   0.00004     &    0.00060    $\pm$   0.00002     &    0.00067    $\pm$   0.00003     &    0.00095    $\pm$   0.00003     &    0.045    $\pm$   0.001     \\
   0.35       $\pm$   0.01        &    0.69       $\pm$   0.01        &    1.51       $\pm$   0.03        &    0.72       $\pm$   0.09        &    0.49       $\pm$   0.01        &    \nodata \\
\hline
    2                             & 131.28860                         & -43.63342                         &   0.00016   $\pm$   0.00001       &   0.00025   $\pm$   0.00005       &   0.0027    $\pm$   0.0003      \\
   0.0165     $\pm$   0.0013      &    0.000044   $\pm$   0.000003    &    0.00014    $\pm$   0.00001     &                        \nodata    &    0.000091   $\pm$   0.000020    &    0.0171     $\pm$   0.0007   \\
   0.63       $\pm$   0.02        &    1.09       $\pm$   0.02        &    2.70       $\pm$   0.03        &    3.08       $\pm$   0.04        &    1.63       $\pm$   0.03        &    \nodata \\
\hline
    3                             & 131.39476                         & -43.85289                         &                        \nodata    &                        \nodata    &                       \nodata   \\
                       \nodata    &    0.00175    $\pm$   0.00008     &    0.00166    $\pm$   0.00005     &    0.0041     $\pm$   0.0002      &    0.0091     $\pm$   0.0007      &                       \nodata  \\
\enddata
\tablenotetext{a}{Coordinates are given in degrees; fluxes ($F$) and their uncertainties ($\Delta F$) are given in Jy.}
\tablenotetext{b}{The uncertainty of the 1.2~mm flux is 20\% \citep{Mas07}.}
\end{deluxetable}

% Tab common protostars
%%%% NUOVO 16 ago 14:
% 2800>-- CII --> 950 >-- FS --> 350 >-- CI --> 70 >-- C0
\begin{deluxetable}{lccccc}
\tablecaption{Protostars detected by both {\it Spitzer} and {\it Herschel} \label{tab_common_protostars}}
\tablewidth{0pt}
\tablehead{ \colhead{NAME} & \colhead{RA} & \colhead{DEC} & \multicolumn{2}{c}{Classification}          & \colhead{$\Delta \Theta$\tablenotemark{a}} \\
         {\it Spitzer}-PSC &  (deg)       & (deg)         & \colhead{MIR\tablenotemark{b}}  & \colhead{FIR\tablenotemark{c}} & \colhead{\arcsec}  } 
\startdata                                        % slope    &  Tbol 
    8036   & 131.28840 & -43.63255   & I    &    0    &          3.1  \\ % 2.2      &  41.7 
    23011  & 131.44509 & -43.38967   & I    &    I    &          0.7  \\ % 1.3      &  77.2 
    41657  & 131.63487 & -43.54801   & I    &    0    &          6.0  \\ % 2.7      &  55.6 
    42629  & 131.64522 & -43.90839   & I    &    I    &          6.9  \\ % 2.7      & 143.8 
    58569  & 131.80015 & -43.37957   & I    &    I    &          0.6  \\ % 0.65     &  91.1 
    65153  & 131.85542 & -43.81557   & I    &    I    &          0.6  \\ % 1.0      & 238.6 
    68231  & 131.88064 & -43.89669   & I    &    0    &          3.9  \\ % 0.43     &  61.1 
    124358 & 132.33936 & -44.03091   & I    &    0    &          6.4  \\ % 0.73     &  45.8 
    19667  & 131.41093 & -43.91746   & FS   &    I    &          1.4  \\ % 0.2      & 259.4 
    59696  & 131.80945 & -43.30581   & FS   &    FS   &          4.7  \\ % 0.1      & 566.6 
\enddata
\tablenotetext{a}{Distance between the positions reported in the MIR and FIR catalog.}
\tablenotetext{b}{Based on the slope in the 2--24~$\mu$m spectral range.}
\tablenotetext{c}{Based on the value of the bolometric temperature (see text).}
\end{deluxetable}

% Tab dei conteggi 
\begin{deluxetable}{lcccccccccc}
\tabletypesize{\scriptsize} 
%\rotate
\tablecaption{Comparison of the {\it Spitzer} YSOs in VMR-D with other SFRs \label{tab_comparison}}
\tablewidth{0pt}
\tablehead{ \colhead{Region} & \colhead{Area}      & \colhead{Dist} & \colhead{YSO}       & \colhead{YSO}          & \colhead{YSO}          & \colhead{Class I}& \colhead{Flat}& \colhead{Class II}& \colhead{Class III}\\ 
            \colhead{}       & \colhead{(deg$^2$)} & \colhead{(pc)} & \colhead{}          & \colhead{(deg$^{-2}$)} & \colhead{(pc$^{-2}$)}  & \colhead{}       & \colhead{}    & \colhead{}        & \colhead{}          } 
\startdata
   this work      & 0.54     &  700  &    499      &   925  &    6.1  &    82 (17\%)   & 52 (10\%)   & 129 (26\%)   & 236 (47\%)    \\
   Paper I        & ''       &  ''   &    297      &   522  &    3.5  &    39 (13.2\%) & 51 (17.3\%) & 142 (47.6\%) & 65 (21.9\%)   \\
\hline
   Cha II         & 1.04     &  178  &     26      &    25  &    2.6  &     2 (8\%)    &  1 (4\%)    & 19 (73\%)    & 4  (15\%)     \\
   Lupus          & 3.10     &  150  &     94      &    30  &    3.3  &     5 (5\%)    & 10 (11\%)   & 52 (55\%)    & 27 (29\%)     \\
   Perseus        & 3.86     &  250  &    385      &   100  &    5.2  &    87 (23\%)   & 42 (11\%)   & 225 (58\%)   & 31 (8\%)      \\
   Serpens        & 0.85     &  260  &    227      &   267  &    13   &    36 (16\%)   & 23 (10\%)   & 140 (62\%)   & 28 (12\%)     \\
   Ophiucus       & 6.60     &  125  &    292      &    44  &    9.3  &    35 (12\%)   & 47 (16\%)   & 176 (60\%)   & 34 (12\%)    
\enddata
\end{deluxetable}

% Tab delle caratteristiche 
\begin{deluxetable}{lccc}
\tabletypesize{\scriptsize} 
%\rotate
\tablecaption{Comparison of SF rate, efficiency and depletion time \label{tab_SF_rate}}
\tablewidth{0pt}
\tablehead{ \colhead{Region} & \colhead{SFR/Area}                        & \colhead{SF efficiency}            & \colhead{depletion} \\ 
            \colhead{}       & \colhead{(M$_\sun$ Myr$^{-1}$ pc$^{-2}$)} & \colhead{(M$_\mathrm{YSO}$/(M$_\mathrm{cl}$+M$_\mathrm{YSO}$))} & \colhead{(Myr)}     } 
\startdata
   this work                & 2.3   & 0.014  & 96 \\
%   Paper I                 & 2.6   & 0.02--0.03 & 45--70   \\ % Pap. I usato: <M(*)>=1Msun
   Paper I\tablenotemark{a} & 1.3   & 0.01--0.02 & 90--140   \\
\hline
   Cha II                   & 0.65  & 0.030  & 66 \\
   Lupus                    & 0.83  & 0.054  & 35 \\
   Perseus                  & 1.3   & 0.038  & 50 \\
   Serpens                  & 3.2   & 0.053  & 35 \\
   Ophiucus                 & 2.3   & 0.063  & 30
\enddata
\tablenotetext{a}{Because the assumed mean mass was 1~M$_\sun$ in Paper I, here the values are recalculated for 0.5~M$_\sun$ (se text).}
\end{deluxetable}

\begin{deluxetable}{lcccccc}
\tabletypesize{\scriptsize} 
%\rotate
\tablecaption{PLF statistics in log(L/L$_\sun$) \label{tab_statistics}}
\tablewidth{0pt}
\tablehead{ \colhead{Region} & \colhead{N}  & \colhead{Mean\tablenotemark{a}} & \colhead{Median\tablenotemark{a}} & \colhead{$\sigma$\tablenotemark{a}} & Completeness(M$_\sun)$ & Reference }
\startdata
            VMR-D            & 53           &  0.31          &   0.25           &   0.58        &  0.35         & This work    \\
            6 low-mass       & 122          & -0.18(-0.09)   &  -0.20(-0.13)    &   0.72(0.78)  &  0.05--0.01   & \citet{Kry12} \\
            3 high-mass      & 425          &  0.15(0.28)    &   0.04(0.18)     &   0.76(0.77)  &  0.05--0.03   & \citet{Kry12} \\
           18 low-mass       & 230          & -0.03(0.07)    &  -0.03(0.11)     &   0.75(0.73)  &  0.2          & \citet{Dun13} \\
\enddata
\tablenotetext{a}{Numbers in parenthesis are obtained after de-reddening.}
\end{deluxetable}

%% FIGURES:
%% the file, uncomment the \plotone, \plottwo, and
%% If you need a layout that cannot be achieved with \plotone or
%% \plottwo, you can invoke the graphicx package directly with the
%% \includegraphics command or use \plotfiddle. For more information,
%% please see the tutorial on "Using Electronic Art with AASTeX" in the
%% documentation section at the AASTeX Web site,
%% http://www.journals.uchicago.edu/AAS/AASTeX.
%%
%% The examples below also include sample markup for submission of
%% supplemental electronic materials. As always, be sure to check
%% the instructions to authors for the journal you are submitting to
%% for specific submissions guidelines as they vary from

%% This example uses \plotone to include an EPS file scaled to
%% 80% of its natural size with \epsscale. Its caption
%% has been written to indicate that additional figure parts will be
%% available in the electronic journal.

% \begin{figure}
% \epsscale{1.0}
% \plotone{fig_LbolK_Tbol_BAKY_alt.eps}
% \caption{ Bolometric Luminosity-Temperature diagram obtained including 10\% of the BLAST and Akari/FIS fluxes (see text). 
%           Five out of 39 sources fall on the Class 0 region: they are 72151, 102091, 106972, 108306, 145123, and their 
%           position in the cloud isreported in Figure~\ref{fig_ds9_sources+coverage}.
%           \label{LbolK_Tbol_alt}}
% \end{figure}

%% The following command ends your manuscript. LaTeX will ignore any text

\end{document}